\documentclass[12pt,a4paper]{article}

\usepackage{epsf,epsfig,psfrag}

\usepackage{geometry}
\usepackage{cite}

\def\beq{\begin{eqnarray}}
\def\eeq{\end{eqnarray}}

\def\be{\begin{equation}}
\def\ee{\end{equation}}

\def\slash#1{#1 \hskip-0.50em /}
\def\Slash#1{#1 \hskip-0.65em /}

\begin{document}

 \newcommand{\bea}{\begin{eqnarray}}
 \newcommand{\eea}{\end{eqnarray}}
 \newcommand{\nn}{\nonumber}
 \newcommand{\dd}{\displaystyle}
 \newcommand{\bra}[1]{\left\langle #1 \right|}
 \newcommand{\ket}[1]{\left| #1 \right\rangle}
 \newcommand{\spur}[1]{\not\! #1 \,}
%---------------------------------------------------

\thispagestyle{empty}

\begin{flushright}
  CERN-PH-TH/2007-167\\
  SLAC-PUB-12823\\
  SI-HEP-2007-14\\
  BARI-TH/07-582
\end{flushright}

\vspace{\baselineskip}

\begin{center}

{\bf \Large
SCET sum rules for \\[0.3em]
$B \to P$ and $B \to V$ transition form factors}

\vspace{3.5\baselineskip}
{\large
  Fulvia De Fazio$^a$,
  Thorsten Feldmann$^{b}$,
  and
  Tobias Hurth$^{c,d,}$\footnote{Heisenberg Fellow}}
\\
\vspace{2em}

{  \it
$^a$ Istituto Nazionale di Fisica Nucleare, Sezione di Bari, Italy}
\\[0.3em]
{  \it
$^b$ Fachbereich Physik,
  Universit\"at Siegen, D-57068 Siegen, Germany}
\\[0.3em]
{  \it
$^c$ CERN, Dept.\ of Physics, Theory Division, CH-1211 Geneva 23, Switzerland}
\\[0.3em]
{ \it $^d$ SLAC, Stanford University, Stanford, CA 94309, USA}

\vspace{3\baselineskip}

\textbf{Abstract}
\vspace{1em}

\parbox{0.95\textwidth}
{
We investigate sum rules for heavy-to-light transition form factors
at large recoil derived from correlation functions with interpolating
currents for light pseudoscalar or vector fields in soft-collinear
effective theory (SCET). We consider both, factorizable and non-factorizable
contributions at leading power in the $\Lambda/m_b$ expansion and to
first order in the strong coupling constant $\alpha_s$, neglecting
contributions from 3-particle distribution amplitudes in the
$B$\/-meson. We pay particular attention to various sources of
parametric and systematic uncertainties. We also discuss
certain form factor ratios where part of the hadronic uncertainties
related to the $B$\/-meson distribution amplitude and to
logarithmically enhanced $\alpha_s$ corrections cancel.
}

\end{center}

\clearpage
\setcounter{page}{1}

\section{Introduction}
\label{sec:intro}

The phenomenological analysis of semi-leptonic and non-leptonic
$B$\/-meson decays into light mesons needs non-perturbative
hadronic input. The theoretically simplest objects are the
heavy-to-light transition form factors. At large recoil energy ($E
\sim m_B/2$) and to leading power in the $\Lambda/m_B$ expansion,
they fulfill a factorization theorem \cite{Beneke:2000wa} which
can be proven using effective-field-theory methods
\cite{Beneke:2003pa,Lange:2003pk} (see also \cite{Bauer:2002aj}):
 \be
\langle M(E)| \bar \psi \, \Gamma_i \, b |B(v)\rangle
=  C_i^{\rm I}(\mu,2E) \, \xi_M(2E,\mu)
+
   C_i^{\rm II}(\mu,2E) \, \frac{\alpha_s C_F}{4\pi} \, \Delta F_M(2E,\mu)
+
   \ldots,
\label{factorization}
\ee
where $M=P,V_\parallel,V_\perp$. The factorization theorem contains
a so-called ``soft'' (non-factorizable) function $\xi_M$ and a
factorizable piece $\Delta F_M$ which is determined by a convolution
of a spectator-scattering kernel $T$ and process-independent
light-cone distribution amplitudes (LCDAs) $\phi_B^+(\omega)$ and
$\phi_M(u)$ for heavy and light mesons, respectively,
\beq
\Delta F_M(2E,\mu) &=&
   T(u,2E,\omega,\mu) \otimes \phi_B^+(\omega,\mu) \otimes
         \phi_M(u,\mu) \,.
\eeq
Finally, the perturbative coefficient functions $C_i^{\rm I,II}$
contain the radiative corrections from short-distance
quantum fluctuations at the scale $\mu \sim m_b$.

The modelling of the soft form-factor terms $\xi_M$ requires
non-perturbative methods. Because of the {\em qualitatively
different}\/ dynamics parametrized by the functions $\xi_M$ and
$\Delta F_M$, it is desirable to follow a non-perturbative
approach where the two contributions can clearly be separated from
the very beginning of the calculation. In previous work
\cite{DeFazio:2005dx,Hurth:2005mm} we have shown that $\xi_M$ and
$\Delta F_M$ can be determined independently via sum rules
derived from correlation functions in
soft-collinear effective theory (SCET
\cite{Bauer:2000yr,Beneke:2002ph}), where the light mesons are
replaced by suitably chosen interpolating currents. Both, the correlation
function for $\xi_M$ and for $\Delta F_M$ involve the light-cone
distribution amplitudes of the $B$\/-meson. Radiative corrections
to the correlation function can be systematically calculated
in perturbation theory, separating  the dynamics at the
intermediate (``hard-collinear'') scale of order
$\sqrt{m_b \Lambda_{\rm QCD}}$ from the soft hadronic
binding effects in the $B$\/-meson. It is to be stressed that
the correlation functions unambiguously factorize into
perturbative short-distance functions and process-independent
LCDAs (the factorization to order $\alpha_s$ accuracy and
neglecting 3-particle LCDAs will be verified explicitly below).
In contrast, the sum rules derived from these correlators
introduce additional sensitivity to non-perturbative
parameters, which not necessarily needs to be
process-independent and thus reflects some irreducible theoretical
uncertainty.

In our previous article we have concentrated on the $B\to \pi$
form factor. In this work we extend and generalize the
discussion to also include transitions to light vector
mesons, $V = \rho, K^*, \ldots$ \ We find it particularly
useful to consider form factor ratios, where the leading
dependence on the hadronic input parameters related to the
$B$\/-meson drops out to some extent.

The paper is organized as follows: In section~\ref{nonfactFF} we
derive the SCET sum rules for the $B \to V_\parallel$, $V_\perp$
non factorizable form factors at ${\cal O}(\alpha_s)$. The  sum
rules for the corresponding factorizable form factors are obtained
in section~\ref{factFF}. Numerical analyses are performed in
section~\ref{numanalysis}, where also the case of $B \to P$ form
factors is reconsidered. The last section is devoted to
conclusions. Some useful formulas are collected in the appendix.

\section{SCET sum rule for non-factorizable form
factor}\label{nonfactFF}

In the following, we will use light-cone variables defined in
terms of two light-like vectors $n_+^2=n_-^2=0$ which are
normalized as $n_+ n_- =2$, such that any vector is
decomposed as
\beq
  a^\mu &=& (n_+a) \, \frac{n_-^\mu}{2} + (n_-a) \, \frac{n_+^\mu}{2} +
  a_\perp^\mu \,.
\eeq We consider a reference frame where the $B$\/-meson velocity
vector satisfies $v_\perp=0$ and $n_+v = n_-v = 1$, and the
final-state momentum $p'$ is purely longitudinal, $p_\perp'=0$. In
this frame the two independent kinematic variables appearing in
the SCET correlation functions (see below) are taken as \beq
 (n_+ p'{}) \simeq 2 E = {\cal O}(m_b) \, , \qquad
 0 > (n_- p'{}) = {\cal O}(\Lambda)\,.
\eeq The dispersive analysis will be performed with respect to
$(n_-p')$ for fixed values of $(n_+p')$.

In SCET the non-factorizable
form factors for transitions between a $B$~meson and
a light vector meson are defined in terms of matrix elements of
the two current operators
\cite{Beneke:2003pa}
\beq
 J_0^\parallel &=& \bar \xi_{\rm hc} W_{\rm hc}
  \, (-\gamma_5) \, Y_s^\dagger h_v \,, \\
 J_0^{\nu_\perp} &=& \bar \xi_{\rm hc} W_{\rm hc}
  \, (\gamma^{\nu_\perp}) \, Y_s^\dagger h_v\,,
\label{eq:J0}
\eeq
where $\xi_{\rm hc}$ is the ``good'' light-cone component of the
light-quark spinor with $\slash n_- \xi_{\rm hc} =0$,
and  $h_v$ is the usual HQET field.
The hard-collinear and soft Wilson lines,
$W_{\rm hc}$ and $Y_s^\dagger$,
appear to render the form-factor definitions
manifestly gauge invariant in SCET.
The normalization conventions
for the corresponding two form factors $\xi_\parallel$ and $\xi_\perp$
are as in \cite{Beneke:2000wa}
\beq
\langle V(p',\varepsilon) | J_0^\parallel | \bar B(v)\rangle
   &=&
 \frac{(n_+ \varepsilon^*)}{2} \, (n_+p') \, \xi_\parallel(n_+p')
\,,\\
  \langle V(p',\varepsilon) | J_0^{\nu_\perp} | \bar B(v)\rangle
   &=& \frac{i}{2} \, (n_+ p') \, \xi_\perp(n_+ p') \,
    \epsilon^{\nu_\perp \kappa_\perp \sigma \tau} \varepsilon_{\kappa_\perp}^*
   n_{-\tau} n_{+\sigma}
\eeq
where $\epsilon^{0123}=+1$.

The SCET sum rules are derived from a dispersive
analysis of the correlation functions
\beq
  \Pi_\parallel(n_-p') &=& i \int d^4x \,
   e^{i p'{} x} \, \langle 0 | T[ J_V^\parallel(x) J_0^\parallel(0)]
     | B(v) \rangle\,,
\label{eq:Pipar}
\\
\frac12\,
 \epsilon^{\mu_\perp \nu_\perp \sigma \tau}
 \, n_{+\sigma} \, n_{-\tau} \,
 \Pi_\perp(n_-p') &=& i \int d^4x \,
   e^{i p'{} x} \, \langle 0 | T[ J_V^{\mu_\perp}(x) J_0^{\nu_\perp}(0)]
     | B(v) \rangle\,,
\label{eq:Piperp}
\eeq
where the longitudinal and transverse polarization-state of the
light vector meson is replaced by the interpolating current
\beq
  J_V^\parallel(x)
  & = & - i \, \bar \xi_{\rm hc}(x)  \,
          \slash n_+  \, \xi_{\rm hc}(x) - i
\left( \bar \xi_{\rm hc} W_{\rm hc}(x)
       \, \slash n_+ \, Y_s^\dagger q_s(x) + {\rm h.c.}\right) \,,
\label{eq:Jpar}
\eeq
and
\beq
  i J_V^{\mu_\perp}(x)
  & = & \bar \xi_{\rm hc}(x)  \,
          i \slash n_+ \gamma^{\mu_\perp} \, \xi_{\rm hc}(x) +
       \left( \bar \xi_{\rm hc} W_{\rm hc}(x)
       \, i \slash n_+ \gamma^{\mu_\perp} \, Y_s^\dagger q_s(x) + {\rm h.c.}\right) \,,
\label{eq:Jperp}
\eeq
respectively.
Here we denoted soft quark fields in SCET as $q_s$.
Notice that soft-collinear interactions require a multi-pole expansion
of soft fields \cite{Beneke:2002ph} which is always understood implicitly.
The matrix element of the interpolating vector-meson currents
are given as
\beq
  \langle 0| J_V^\parallel | V(p',\varepsilon)\rangle
  &=& m_V \, (n_+ \varepsilon) \, f_V^\parallel \, ,
\\
  \langle 0| i J_V^{\mu_\perp} | V(p',\varepsilon)\rangle
  &=& (n_+ p') \, \varepsilon^{\mu_\perp} \, f_V^\perp(\mu) \,.
\eeq The dispersion relation for the above correlation functions,
after Borel transform with respect to the variable $n_-p'{}$,
reads \beq
  \hat B\left[\Pi_{\parallel,\perp} \right](\omega_M)
&=&  \int\limits_0^\infty d\omega' \, \frac{1}{\omega_M}
     \, e^{-\omega'/\omega_M} \,  \frac{1}{\pi} \,
     {\rm Im}\left[\Pi_{\parallel,\perp}(\omega')\right]\,,
\label{eq:Borel}
\eeq
where $\omega_M \equiv M^2/(n_+p')$ is the Borel parameter.
The hadronic representation of the same correlation function
reads
\begin{equation}
\Pi_{\parallel,\perp}^{\rm HAD}(n_-p')
= \Pi_{\parallel,\perp}(n_- p') \Big|_{\rm res.}+ \Pi_{\parallel,\perp}(n_- p')
\Big|_{\rm cont.} \,,
\label{eq:hadside}
\end{equation}
where the first term represents
the contribution of the $V$ meson, while the second takes into account
the role of higher states and continuum above an effective
threshold $\omega_s \equiv s_0/(n_+p')$. Assuming quark-hadron duality implies
$\Pi_{\parallel,\perp}(n_-p')=\Pi_{\parallel,\perp}^{\rm HAD}(n_-p')$.
For the longitudinal part we obtain the
resonance contribution
\beq
  \Pi_\parallel(n_- p') \Big|_{\rm res.} &= &
   \frac{\langle 0|J_V^\parallel |V(p',\varepsilon)\rangle
         \langle V(p',\varepsilon)|J_0^\parallel|B(v)\rangle}
    {m_V^2-p'{}^2}
 \ = \
\frac{n_+p'}{2m_V} \,
\frac{ (n_+p')^2 \, \xi_\parallel(n_+p') \, f_V^\parallel}
{m_V^2-p'{}^2}\,.
\eeq
Apart from an overall factor $(n_+p')/2m_V = E_V/m_V$
this has the same form as the expression for $B \to \pi\,$ if one
replaces $(f_\pi,\xi_\pi) \to (f_V^\parallel,\xi_V^\parallel)$.
However, now we should not neglect the vector-meson mass in the
hadronic side of the sum rule. As a consequence, the Borel
transform of the resonance contribution reads
\begin{equation}
\hat B\left[\Pi_\parallel \right](\omega_M)\Big|_{\rm res.} =
\frac{n_+p'}{2m_V} \, \exp\left[-\frac{m_V^2}{(n_+p')\, \omega_M}\right]
{(n_+p') \, \xi_\parallel(n_+ p') \, f_V^\parallel \over \omega_M} \,.
\end{equation}
Modelling the continuum by the perturbative result and subtracting
it on both sides of (\ref{eq:hadside}), we obtain the sum rule
\beq
 \frac{n_+p'}{2m_V} \, \xi_\parallel(n_+p') &= &
   \frac{1}{f_V^\parallel \, (n_+p')} \,
   \exp\left[\frac{m_V^2}{(n_+p')\, \omega_M}\right]
   \, \int\limits_0^{\omega_s} d\omega' \, e^{-\omega'/\omega_M} \, {1 \over
   \pi} {\rm Im}[\Pi_\parallel(\omega')]
 \,.
\label{eq:sumpar}
\eeq
Analogously, the sum rule for the transverse form factor follows
as
\beq
 \xi_\perp(n_+p') &= &
   \frac{1}{f_V^\perp \, (n_+p')} \,
   \exp\left[\frac{m_V^2}{(n_+p')\, \omega_M}\right]
   \, \int\limits_0^{\omega_s} d\omega' \, e^{-\omega'/\omega_M} \, {1 \over
   \pi} {\rm Im}[\Pi_\perp(\omega')]
 \,.
\label{eq:sumperp}
\eeq

%%%%%%%%%%%%%%%%%%%%%%%%%%%%%%%%%%%%%%%%%%%%%%%%%%%%%%%%%%%%%%%%%%%%%%%%%

\subsection{Tree-level result}

At leading power, the tree-level result
for the correlation function involves one
hard-collinear quark propagator, which reads
\beq
  S_F^{\rm hc} = \frac{i}{n_- p'{} - \omega + i\eta}
  \, \frac{\slash n_-}{2}\,,
\eeq
where $\omega = n_- k$, and $k^\mu$ is the momentum of the soft light quark
that  will end up as the spectator quark in the $B$ meson.
The remaining matrix element defines
light-cone distribution amplitudes for the $B$\/ meson in HQET
\cite{Grozin:1996pq,Beneke:2000wa} (for the definition,
see appendix~\ref{app:phiB}).
This leads to
\beq
  \Pi_\parallel (n_- p'{}) &=& \Pi_\perp(n_-p') =
   f_B m_B \int\limits_0^\infty d\omega \,
   \frac{\phi_B^-(\omega)}{\omega - n_-p'{}-i\eta} \,
\eeq
which is identical to the analogous result calculated for the
$B \to \pi$ case \cite{DeFazio:2005dx}.
Inserting the imaginary part into the sum rules
(\ref{eq:sumpar},\ref{eq:sumperp}), the tree-level result for
the soft form factors follows as
\beq
\hat \xi_\parallel(n_+p')\equiv \frac{n_+p'}{2m_V} \,
  \xi_\parallel(n_+p') &= & \frac{f_B m_B}{f_V^\parallel \, (n_+p')} \,
  \exp\left[\frac{m_V^2}{(n_+p')\, \omega_M}\right]
   \int\limits_0^{\omega_s} d\omega' \, e^{-\omega'/\omega_M} \, \phi_B^-(\omega')
 \quad \mbox{(tree level)}\,,
\cr &&
\label{sumpartree}
\eeq
and
\beq
 \xi_\perp(n_+p') &= & \frac{f_B m_B}{f_V^\perp  \, (n_+p')} \,
  \exp\left[\frac{m_V^2}{(n_+p')\, \omega_M}\right]
   \int\limits_0^{\omega_s} d\omega' \, e^{-\omega'/\omega_M} \, \phi_B^-(\omega')
 \quad \mbox{(tree level)}\,.
\cr &&
\label{sumperptree}
\eeq
Notice that the suppression of the longitudinal form factor $\xi_\parallel$
by $m_V/(n_+p')$ compared to $\xi_\perp$ is expected on
general grounds \cite{Charles:1998dr,Beneke:2000wa}.

%%%%%%%%%%%%%%%%%%%%%%%%%%%%%%%%%%%%%%%%%%%%%%%%%%%%%%%%%%%%%%%

\subsection{Radiative corrections from hard-collinear loops}

\begin{figure}[hbt]
\begin{center}
 \psfrag{p}{\scriptsize ${\parallel,\perp}$}
 \psfig{file = 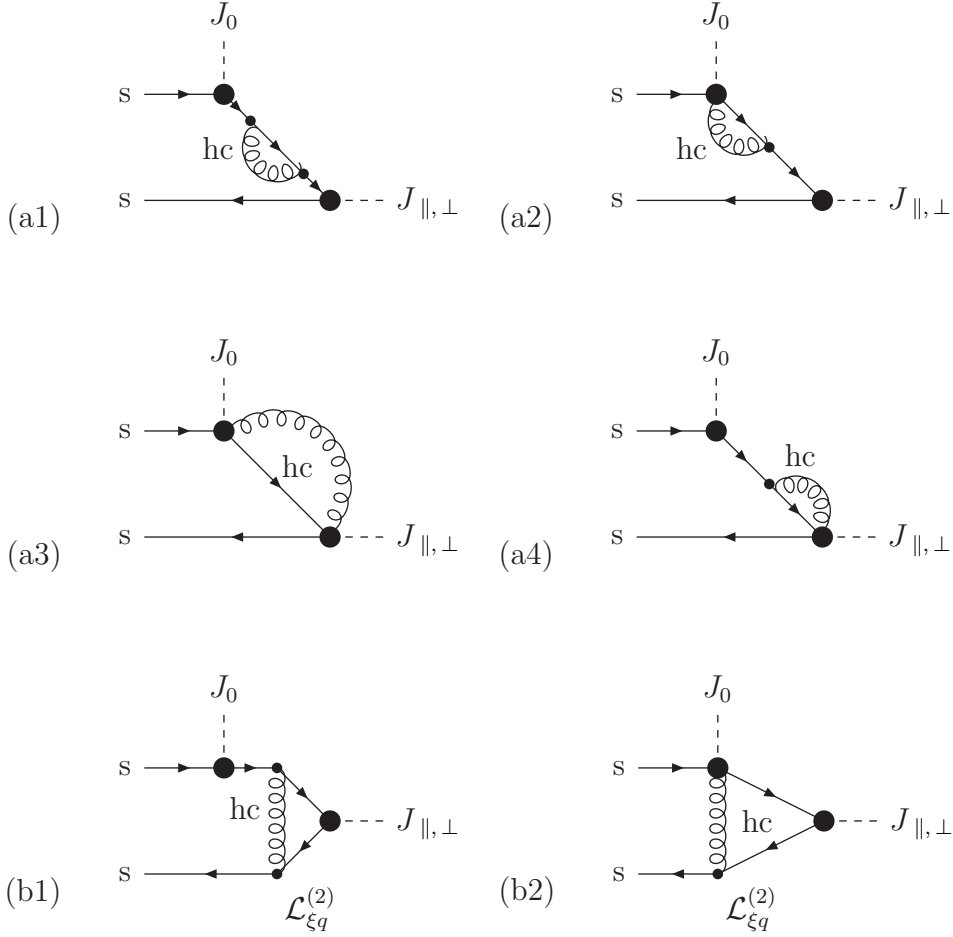}
\end{center}
\centerline{\parbox{0.95\textwidth}{
\caption{\small Diagrams contributing to the sum rule for $\xi_{\parallel
,\perp}$ to order $\alpha_s$ with hard-collinear loops and no
external soft gluons. Diagrams (a2-a4) and (b2) vanish in
light-cone gauge $n_+ A_{\rm hc} =0$. Diagram (a3) vanishes both
in light-cone and in Feynman gauge.} \label{fig:hcloops1}}}
\end{figure}

In SCET short-distance radiative corrections to
the correlation functions (\ref{eq:Pipar},\ref{eq:Piperp})
are represented by hard-collinear loops,
as shown in Fig.~\ref{fig:hcloops1} for the leading
order in $\alpha_s$ (1-loop contributions from three-particle
distribution amplitudes in the $B$\/-meson are neglected
in this work).
The diagrams denoted by (a1-a4) and (b1-b2)
form gauge-invariant subsets, such that the result for either of
the correlation functions can be written as
\beq
   \Pi_{\parallel,\perp}(\omega',\mu)
&=&  f_B(\mu) m_B \int\limits_0^\infty
  \frac{d\omega}{\omega - \omega'{}-i\eta} \, \phi_B^-(\omega,\mu)
  \left\{ 1 + \frac{\alpha_s C_F}{4\pi}
  \left( \mbox{(a1-a4)} + \mbox{(b1-b2)} \right) \right\}_{\parallel,\perp} \,.
\cr &&
\label{hccontri}
\eeq
The corrections from diagrams (a1-a4) are identical for pseudoscalar,
longitudinal and transverse vector mesons and read
\cite{DeFazio:2005dx}
\beq
  \mbox{(a1-a4)} &=&
                  \frac{4}{\epsilon^2} + \frac{3 + 4 L(\mu)}{\epsilon}
  + L(\mu) \, (3 + 2 L(\mu)) + 7 - \frac{\pi^2}{3} \,,
\label{a14}
\eeq
where we have used dimensional regularization in $D=4-2\epsilon$, and
defined the abbreviation
\beq
L(\mu) &=&
\ln \left[-\frac{\mu^2}{(n_+ p')(\omega'-\omega + i\eta)}\right] \,.
\eeq
Notice that the diagrams (a1-a4) also determine the perturbative
corrections to the jet function in inclusive $b\to u$ decays
\cite{Bauer:2003pi,Bosch:2004th} (see also appendix).

For the diagrams (b1-b2) we have to consider soft-collinear
vertices from the sub-leading SCET$_{\rm I}$ Lagrangian ${\cal
L}_{\xi q}^{(2)}$, see \cite{Beneke:2002ph}. Alternatively, one
can single out the hard-collinear integration region from the
corresponding QCD diagrams and use the momentum-space projector
for the $B$\/-meson distribution amplitude, see (\ref{bproj}) in
the appendix. In either case we obtain \beq
  \mbox{(b1-b2)}_\parallel  &=&
-\frac{2}{\epsilon^2}
- \frac{2 L_1 + 2 L(\mu) + 3}{\epsilon}
 \nonumber \\[0.2em]
&& {} -L_1^2-\frac{\omega + 2\omega'}{\omega}\, L_1 - L^2(\mu)-(2 L_1+3) L(\mu)+\frac{\pi^2}{6}-8
\,,
\label{b12par}
\eeq
where we have defined
\beq
  L_1 &=& \ln \left[1- \frac{\omega}{\omega'+i\eta}\right]\,.
\eeq
Again, the expression for $\mbox(b_1-b_2)_\parallel$ coincides with
the $B \to \pi$ case.\footnote{We spotted a calculational error
in our original result for the $B\to \pi$ correlator in \cite{DeFazio:2005dx},
which affects the single-logarithmic terms. The corrected formulas
and numerics will be discussed below.}
For the transverse polarization we obtain
 \beq
  \mbox{(b1-b2)}_\perp  &=&
\mbox{(b1-b2)}_\parallel
-\frac{1}{\epsilon} - L(\mu) -L_1 \,.
\label{b12perp}
\eeq
The extra contribution with $1/\epsilon$
is related to the anomalous dimension
of the tensor current (\ref{eq:Jperp}), and
the corresponding extra $\mu$ dependence
cancels that of $f_V^\perp(\mu)$,
\beq
  \frac{\partial f_V^\perp(\mu)}{\partial \ln \mu^2}
  &=& \frac{\alpha_s C_F}{4\pi} \, f_V^\perp(\mu) + \ldots
\eeq

\subsubsection{Cancellation of factorization-scale dependence}

The physical form factors defined by (\ref{factorization}) have to be
independent of the factorization scale $\mu$, which have been introduced to
separate the momentum regions contributing to the $B\to M$ transition.
In our previous paper \cite{DeFazio:2005dx} we have already verified that
the double-logarithmic dependence on $\mu$, related to the universal cusp
anomalous dimension, drops out when combining the scale-dependence of the
hard matching coefficients between QCD and SCET \cite{Bauer:2000yr},
\beq
 \frac{d}{d \ln \mu} \, C_i(\mu) &=& - \frac{\alpha_s C_F}{4\pi}
 \left( \Gamma_{\rm cusp}^{(1)} \, \ln \frac{\mu}{m_b} + 5 \right) C_i(\mu) + \ldots
\label{Ccontr} \eeq with $\Gamma_{\rm cusp}^{(1)}=4$, and the
explicit scale dependence of the renormalized SCET correlation
function in (\ref{hccontri}), 
\beq && \frac{d}{d\ln\mu} \,
\Pi_{\parallel}(\omega',\mu)   =  - \frac{\alpha_s C_F}{4\pi} \,
 f_B(\mu) m_B \int\limits_0^\infty \frac{d\omega}{\omega-\omega'} \,
 \int_0^\infty d\tilde\omega \,
 \gamma_-^{(1)}(\omega,\tilde\omega,\mu) \, \phi_B^-(\tilde\omega,\mu)
\nonumber\\[0.2em]
&& \quad + \frac{\alpha_s C_F}{4\pi}
  f_B(\mu) m_B \int\limits_0^\infty
  \frac{d\omega}{\omega - \omega'{}}
  \left(
  \Gamma_{\rm cusp}^{(1)} \, L(\mu)
- \Gamma_{\rm cusp}^{(1)} \, L_1
  + 3
\right) \phi_B^-(\omega,\mu) + \ldots
\label{Picontr}
\eeq
Here we have only quoted the result for $\Pi_\parallel$, since $\Pi_\perp$
simply differs by a single log related to the anomalous dimension of the
decay constant $f_\perp(\mu)$ (see above).
Neglecting possible 3-particle contributions,
the scale dependence of the $B$\/-meson LCDA in HQET is described by the
anomalous dimension $\gamma_-$,
\beq \label{eq:NRLCDABEvol}
\frac{d}{d \ln \mu} \; \phi_B^-(\omega;\mu) & = & - \, \frac{\alpha_s C_F}{4\pi}
\int_0^\infty d\tilde\omega \; \gamma_-^{(1)}(\omega,\tilde\omega;\mu)
 \; \phi_B^-(\tilde\omega;\mu) + \ldots
\label{gamma_conv}
\eeq
which gives rise to the first line in (\ref{Picontr}).
The second line arises from the
explicit scale-dependence induced by the NLO
hard-collinear $\alpha_s$ corrections to the correlator,
where we also took into account the anomalous
dimension of the $B$\/-meson decay constant in HQET,
$df_B/d\ln\mu = 3 \, \frac{\alpha_s C_F}{4\pi}+\ldots$

While the anomalous dimension for the LCDA $\phi_B^+(\omega,\mu)$
has been known for some time \cite{Lange:2003ff}, the anomalous
dimension for the LCDA $\phi_B^-(\omega,\mu)$ has  been calculated
only recently \cite{Bell:2007} with the result \beq
  \gamma_-^{(1)}(\omega,\tilde\omega;\mu) &=&
\left( \Gamma_{\rm cusp}^{(1)} \, \ln \frac{\mu}{\omega} - 2
\right) \delta(\omega-\tilde\omega)
- \Gamma_{\rm cusp}^{(1)} \, \frac{\theta(\tilde\omega-\omega)}{\tilde\omega}
\nonumber \\[0.2em]
& & {} - \Gamma_{\rm cusp}^{(1)} \, \omega \left[\frac{\theta(\tilde\omega-\omega)}{\tilde\omega(\tilde\omega-\omega)} \right]_+
- \Gamma_{\rm cusp}^{(1)} \, \omega \left[
\frac{\theta(\omega-\tilde\omega)}{\omega(\omega-\tilde\omega)} \right]_+ \,.
\label{gammam}
\eeq
This enables us to also verify the perturbative cancellation of the single
logarithmic terms.
For that purpose, we have to insert the convolution (\ref{gamma_conv})
into the tree-level term in (\ref{hccontri}). Using integration-by-parts,
we find
\beq
&& {} - \int_0^\infty \, \frac{d\omega}{\omega-\omega'}
 \, \int_0^\infty d\tilde\omega \, \gamma_-^{(1)}(\omega,\tilde\omega,\mu) \, \phi_B^-(\tilde\omega,\mu)
\nonumber \\[0.2em]
&=&
{}
\int_0^\infty \, \frac{d\omega}{\omega-\omega'} \left(-\Gamma_{\rm cusp}^{(1)} \, \ln \frac{\mu}{\omega-\omega'} + \Gamma_{\rm cusp}^{(1)}\, L_1 + 2  \right) \phi_B^-(\omega,\mu)
\,.
\label{phicontr}
\eeq
Inserting (\ref{phicontr}) into (\ref{Picontr}) and combining with (\ref{Ccontr}),
we find indeed complete cancellation of all $\mu$\/-dependent terms,
which proofs the factorization of the SCET correlator to
order $\alpha_s$ accuracy (in the absence of 3-particle LCDAs).

\subsubsection{Contribution to the sum rule}

In \cite{DeFazio:2005dx} we have calculated the imaginary part of
the correlation function using the approximation $\phi_B^-(\omega)
\simeq \phi_B^-(0)$ which is valid for $\omega_{M,s}\ll \Lambda$.
Using the relations provided in the appendix, it is possible to
derive the exact expressions which, inserted in
Eq.~(\ref{eq:Borel}), give the final result for the
Borel-transformed and continuum-subtracted correlator:
 \beq \hat B
[\Pi_{\parallel,\perp}](\omega_M)- \mbox{cont.} &=&
 \frac{ f_B m_B }{\omega_M}
 \, \int\limits_0^{\omega_s} d\omega' \, e^{-\omega'/\omega_M} \,
\left\{
  - \int_{\omega'}^\infty d\omega \, f_{\parallel,\perp}(\omega,\omega',\mu) \, \frac{d\phi_B^-(\omega,\mu)}{d\omega}
\right.
\nonumber \\[0.2em]
&& \left. \quad {}
  + \int_0^{\omega'} d\omega
\left[
\frac{g_{\parallel,\perp}(\omega,\omega',\mu)}{\omega-\omega'}
\right]_+
\phi_B^-(\omega,\mu) \right\}
\label{pifinal}
\eeq
where we introduced the functions $f_{\parallel,\perp}=1 + {\cal
O}(\alpha_s)$ and $g_{\parallel,\perp} %,h_{\parallel,\perp}
= {\cal O}(\alpha_s)$:
\beq
  f_\parallel(\omega,\omega',\mu) &=&
1 + \frac{\alpha_s C_F}{4\pi} \left(
L_0^2 +(1+2L_0) \, \ln \frac{\omega'}{\omega}
- (3+2L_0) \, \ln \left[1 - \frac{\omega'}{\omega} \right]
- 1 + \frac{\pi^2}{6}
\right)
\cr &&
\label{fterm}
\\[0.2em]
f_\perp(\omega,\omega',\mu) &=&
1 + \frac{\alpha_s C_F}{4\pi} \left(
L_0^2-L_0 +(2+2 L_0) \, \ln \frac{\omega'}{\omega}
- (4+2 L_0) \, \ln \left[1 - \frac{\omega'}{\omega} \right]
- 1 + \frac{\pi^2}{6}
\right)
\cr &&
\label{fperpterm}
\eeq
and
\beq
  g_{\parallel}(\omega,\omega',\mu) =
  g_\perp(\omega,\omega',\mu)+\frac{\alpha_s C_F}{4\pi}
&=& \frac{\alpha_s C_F}{4\pi} \left( 2 L_0 - 4 L_1 \right) \,,
\eeq  where
\beq &&  L_0 =  \ln \left[\frac{\mu^2}{(n_+p')
\omega'}\right] \,. \eeq
The final sum rules at ${\cal O}(\alpha_s)$ are obtained
by replacing $\phi_B^-(\omega')$ in (\ref{sumpartree},\ref{sumperptree})
by the curly brackets in (\ref{pifinal}),
\beq
 \phi_B^-(\omega') &\to& \phi_{\parallel,\perp}^{\rm eff}(\omega',n_+p',\mu)
\equiv
\left\{
  - \int_{\omega'}^\infty d\omega \, f_{\parallel,\perp}(\omega,\omega',\mu) \, \frac{d\phi_B^-(\omega,\mu)}{d\omega}
\right.
\nonumber \\[0.2em]
&& \left. \qquad {} \qquad
  + \int_0^{\omega'} d\omega
\left[
\frac{g_{\parallel,\perp}(\omega,\omega',\mu)}{\omega-\omega'}\right]_+
\phi_B^-(\omega,\mu) \right\}
\label{eq:phieff}
\eeq

Notice that the integration variable in the dispersion integral
is restricted to values $\omega' \leq \omega_s = s_0/(n_+p')$,
and therefore the natural choice for the scale $\mu$
in the function $L_0$ is soft, $\mu_{\rm soft}^2 \sim s_0 = {\cal O}(1~{\rm GeV}^2)$
(i.e.\ independent of the heavy-quark mass).
On the other hand, the logarithm $\ln [\omega'/\omega]$
in the function $f(\omega,\omega',\mu)$ is evaluated for values of
$\omega \geq \omega'$,
and therefore the natural scale for $\omega$ is set by the $B$\/-meson LCDA,
$\omega \sim \omega_0 \simeq 0.5~{\rm GeV}$.
The identified large logarithms,
arising in the formal limit $\mu^2 \sim s_0 \ll \omega_0 (n_+p')$,
stem from the spectator diagrams (b1,b2) in Fig.~\ref{fig:hcloops1}
and have the same origin as the endpoint singularities appearing
in the QCD factorization approach.\footnote{A similar situation arises for
heavy-to-light form factors with non-relativistic bound states,
as discussed in \cite{Bell:2005gw}. Here, the finite (constituent)
quark masses provide an intrinsic infrared regulator, such that
the form factors are well-defined in fixed-order perturbation
theory. Still, one encounters large logarithms which cannot be
attributed to the evolution of mesonic light-cone wave function
or hard interaction kernels.}
The appearance of different scales in $\phi_{\rm eff}$ thus
indicates that the resummation of large logarithms
in the heavy-quark limit \emph{for the sum rule} is not complete.

\label{largelog}

A formal solution of this problem within the sum-rule approach
would require to consider two-loop corrections to the correlator and to better
understand the evolution equations of the B-meson LCDA $\phi_B^-(\omega)$
in the presence of 3-particle LCDAs, which
goes beyond the scope of this work. Usually, one takes a more pragmatic point of view
and ignores the formal logarithmic enhancement of the perturbative coefficients,
since numerically the effect does not appear to be very large.
However, one should be aware of the fact that the presence of these
non-factorizable logarithms prevents us from performing the
very heavy-quark limit in the final sum rule for the soft
form factor, see also section~\ref{sec:mQ} below.

\section{SCET sum rule for QCD-factorizable contributions}\label{factFF}

\begin{figure}[hbt]
\begin{center}
 \psfrag{p}{\scriptsize $\parallel,\perp$}
 \psfig{file = 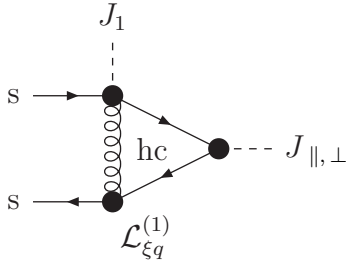}
\end{center}
\centerline{\parbox{0.95\textwidth}{\caption{\small Factorizable contribution to heavy-to-light form factors
 in the SCET sum rule approach.}}}
\label{fig:lead}
\end{figure}

In our previous paper \cite{DeFazio:2005dx}, we have shown that at
leading-order in $\alpha_s$ our sum-rule approach reproduces the
result for the {\em factorizable\,}\/ form-factor contribution from
hard-collinear spectator scattering as derived in
\cite{Beneke:2003pa,Beneke:2000wa} (spectator-scattering corrections
to heavy-to-light form factors at NLO in QCD factorization are also known
by now \cite{Beneke:2005gs,Kirilin:2005xz,Becher:2004kk}). We extend the results derived
in \cite{DeFazio:2005dx} for $B \to \pi$ transition to the case of $B$
decays to a light vector meson.

In case of the pion the corrections to the symmetry relations are
derived by an independent sum rule starting with the correlation
function \beq
  \Pi^1_\pi (p') &=& i \int d^4x \,
   e^{i p'{} x} \, \langle 0 | T[ J_\pi(x) J_1^\pi (0)] | B(p_B) \rangle
   \,,
\label{piscet1}
\eeq
where
\beq
 J_1^\pi  &\equiv &
 \bar \xi_{\rm hc} \,  g \, \slash A^\perp_{\rm hc} \,  h_v \,.
\eeq The analogous function for the $\rho$ with longitudinal
polarization is \beq
  \Pi^1 _\parallel(p') &=& i \int d^4x \,
   e^{i p'{} x} \, \langle 0 | T[ J_\rho^\parallel(x) J_1^\parallel(0)]
     | B(v) \rangle\,,
\eeq
with the factorizable current\footnote{Note that this current could be in principle further reduced to $J_1^\parallel  \equiv i \epsilon^{\mu \sigma}_\perp \,\,\,  \bar \xi_{\rm hc} \,  g \, A^\perp_{{\rm hc}\, \mu} \,\gamma_{\sigma}^\perp   h_v$.}
\beq
 J_1^\parallel  &\equiv &
 \bar \xi_{\rm hc} \,  g \, \slash A^\perp_{\rm hc} \,\gamma_5
 h_v \,.
\eeq In the case of $\rho$ with transverse polarization there are
two possible currents  having a nonvanishing matrix element
between the $B$ meson and a transverse $\rho$ state, which are
\beq
 J_{1,1}^{\nu_\perp}\,  \equiv \,
 \bar \xi_{\rm hc} \,  g \, A^{\nu_\perp}_{\rm hc}   h_v \,\, , \,\,\,
J_{1,5}^{\nu_\perp}\,  \equiv \,
 \bar \xi_{\rm hc} \,  g \, A^{\nu_\perp}_{\rm hc} \,\gamma_5  h_v
 \,.
\eeq As a consequence, we consider two different correlation
functions:
\beq \Pi_\perp^{1,1 \mu_\perp \nu_\perp} &=&    i \int
d^4x \,
   e^{i p'{} x} \, \langle 0 | T[ J_\rho^{\mu_\perp}(x) J_{1,1}^{\nu_\perp}(0)]
     | B(v) \rangle = (-1) \frac12\, \epsilon_\perp^{\nu\mu}  \,
 \Pi_\perp^{1,1}(p')
\eeq
with  $\epsilon^{\nu \mu}_\perp \equiv
\epsilon^{\nu_\perp \mu_\perp \sigma \tau}\, n_{-\tau} \, v_{\sigma}$, or alternatively
\beq
\Pi_\perp^{1,5 \mu_\perp \nu_\perp} &=&    i \int d^4x \,
   e^{i p'{} x} \, \langle 0 | T[ J_\rho^{\mu_\perp}(x) J_{1,5}^{\nu_\perp}(0)]
     | B(v) \rangle  = (-i) \frac12\,
 g^{\mu \nu}_\perp  \, \Pi_\perp^{1,5}(p') \,.
\eeq The leading contributions to those correlation functions are
given by the diagram in Fig.~\ref{fig:lead} which involves the
insertion of one interaction vertex from the order-$\lambda$
soft-collinear Lagrangian
$$
 {\cal L}_{\xi q}^{(1)}
\ = \ \bar \xi_{\rm hc} \, g_s \Slash A_{\rm hc,\perp} \, q_s +
\mbox{h.c.} \,.
$$

As in the pion case, the resulting hard-collinear loop integrals
are  UV- and IR-finite and  all  the scalar functions turn out to
be equal: \beq \Pi^1 (p') \equiv   \Pi^1_\pi  (p') =
\Pi^1_\parallel  (p')   = \Pi^{1,1}_\perp (p') = \Pi^{1,5}_\perp
(p') \,. \eeq
The calculation yields \beq
  \Pi^1 (p') &=& - \frac{\alpha_s C_F}{4\pi}
    (n_+p') \, \int_0^\infty d\omega \,
    \frac{f_B m_B \, \phi_B^+(\omega)}{\omega} \,
    \ln \left[1 - \frac{\omega}{n_-p'+i\eta}\right] \,,
\eeq the imaginary part of which reads as \beq
   \frac{1}{\pi} \, {\rm Im}[\Pi^1(\omega')]
   &=& - \frac{\alpha_s C_F}{4\pi} \,
    (n_+p') \, \int_0^\infty d\omega \,
    \frac{f_B m_B \, \phi_B^+(\omega)}{\omega} \,
    \theta[\omega-\omega'] \,.
\eeq
Inserting this into the dispersion relation analogous
to the one of the tree-level sum rule
\beq
  \Pi^1 (p')
  &=& \frac{1}{\pi} \int_0^\infty d\omega' \,
      \frac{{\rm Im}[\Pi_1(\omega')]}{\omega'-n_-p' - i\eta}\,,
\eeq
and performing the Borel transformation and
subtracting the continuum contribution, we obtain
\beq
   \hat B\left[\Pi^1 \right](\omega_M) \Big|_{\rm res.}
 &=& - \frac{\alpha_s C_F}{4\pi} \,
     (n_+p')
    \int_0^\infty d\omega \,
    \frac{f_B m_B \, \phi_B^+(\omega)}{\omega}\,
\nonumber \\
&& \left(1 - e^{-\omega_s/\omega_M} -
         \theta(\omega_s - \omega)\left( e^{-\omega/\omega_M}
       - e^{-\omega_s/\omega_M} \right)\right).
\eeq
Let us first reconsider the pion case. The hadronic expression for
the  contribution of the lowest lying resonance (the pion) reads
in this case \beq
  \Pi^1_\pi(p') \Big|_{\rm res.} &= &
   \frac{\langle 0|J_\pi |\pi(p')\rangle \langle \pi(p')|J_1^\pi|B(p_B)\rangle}
    {m_\pi^2-p'{}^2}
 \ = \
 \frac{ (n_+p')\, f_\pi\,\langle \pi(p')|J_1^\pi|B(p_B)\rangle}{m_\pi^2-p'{}^2}\,.
\eeq
Thus, its Borel transform is given by
\begin{equation}
\hat B\left[\Pi^1_\pi \right](\omega_M)\Big|_{\rm res.}={f_\pi
\langle \pi |J_1^\pi|B\rangle \over \omega_M} \, e^{-m_\pi^2/(n_+
p' \omega_M)} \,.
\end{equation}
Equating the two expressions leads to the sum rule for the factorizable contribution in the pion case which reduces to the result derived in \cite{DeFazio:2005dx} for $m_\pi = 0$:
\beq
\langle \pi|J_1^\pi|B\rangle &=&   - \frac{\alpha_s C_F}{4\pi} \,
    \frac{(n_+p') f_B m_B \omega_M}{f_\pi} \, e^{m_\pi^2/(n_+ p' \omega_M)}\\
&\times&  \int_0^\infty d\omega \,
    \frac{\phi_B^+(\omega)}{\omega}\, \left(1 - e^{-\omega_s/\omega_M} -
         \theta(\omega_s - \omega)\left( e^{-\omega/\omega_M}
       - e^{-\omega_s/\omega_M} \right)\right) \,.
\eeq

In previous work \cite{Beneke:2000wa}, the factorizable contribution
was expressed in terms
of the quantities $\Delta F_\pi , \Delta F_\parallel$, and $\Delta F_\perp$.
As was explicitly shown in the appendix of \cite{DeFazio:2005dx},
one can identify the matrix element corresponding to the factorizable contribution  as follows
\beq
  \langle \pi | J_1^\pi  | B \rangle &=&  \frac{\alpha_s C_F}{4\pi} \, \Delta F_\pi
   \frac{(- m_B^2)}{2} \,.
\label{DeltaF} \eeq Along the same lines, one derives the
identifications of  the quantities $\Delta F_\parallel$ and
$\Delta F_\perp$ with the matrix elements of the factorizable
currents defined above: \beq
  \langle \pi | J_1^\parallel  | B \rangle &=&  \frac{\alpha_s C_F}{4\pi} \, \Delta F_\parallel \,
   \frac{(-m_B^2)}{4} \, n_+ \epsilon^*  \, \frac{m_\rho}{E}\\
\langle \pi | J_{1,1}^{\nu\perp}  | B \rangle &=&  \frac{\alpha_s C_F}{4\pi} \, \Delta F_\perp \,
   \frac{(-i m_B^2)}{4} \, \epsilon^*_{\kappa\perp} \,\epsilon^{\nu\kappa}_\perp \\
\langle \pi | J_{1,5}^{\nu\perp}  | B \rangle &=&  \frac{\alpha_s C_F}{4\pi} \, \Delta F_\perp \,
   \frac{(m_B^2)}{4} \, \epsilon^{* \nu_\perp}\,\,.
\eeq
Then the hadronic side of the sum rule can be directly expressed in terms of the $\Delta F_X$: Using again $|n_+\epsilon^*|^2 = (n_+p')^2/(m_\rho)^2$ we get
\beq
\hat B\left[\Pi^1_\parallel \right](\omega_M)\Big|_{\rm res.}  &=&
\hat B\left[\frac{f_\rho^\parallel n_+p'}{m_\rho^2 -p'^2}\frac{(-m_B^2)}{2} \frac{\alpha_s C_F}{4\pi} \, \Delta F_\parallel \right]\\
&=& {f_\rho^\parallel  \over \omega_M} \, e^{-m_\rho^2/(n_+ p' \omega_M)}
\frac{(-m_B^2)}{2} \frac{\alpha_s C_F}{4\pi} \, \Delta F_\parallel \,,
\eeq
and  using $\epsilon^{*\mu_\perp} \epsilon^{* \nu_\perp} = - g_\perp^{\mu\nu}$,
we get for the scalar functions in the transverse case,
$\hat B\left[\Pi^{1,1}  \right](\omega_M)\Big|_{\rm res.}$ and
$\hat B\left[\Pi^{1,5}  \right](\omega_M)\Big|_{\rm res.}$,  the same
result as if we replace $f_\rho^\parallel$ and $\Delta F_\parallel$ by
$f_\rho^\perp$ and $\Delta F_\perp$ respectively.

Equating the hadronic and SCET expressions  in the various cases
we end up with  the following
sum rule for the quantities  $\Delta F_X$ where $X = \pi,\rho_\parallel,\rho_\perp$:
\beq
  \Delta F_X(\mu,n_+p') &=&  \frac{2 f_B \omega_M (n_+p')}{m_B f_X}
  \, e^{m_X^2/(n_+p' \omega_M)} \,
\nonumber
\\[0.2em] & \times &
  \int_0^\infty \frac{d\omega}{\omega} \, \phi_B^+(\omega,\mu)
   \left( 1- e^{-\omega_s/\omega_M} \, \theta(\omega-\omega_s)
          - e^{-\omega/\omega_M} \, \theta(\omega_s-\omega)\right)
\label{eq:DeltaF1}
\eeq
(with $\omega_M$ and $\omega_s$ depending on the considered meson $X$).

This result can now be compared with
QCD factorization \cite{Beneke:2003pa,Beneke:2000wa}.
If we insert the leading-order sum rule for the decay constants
\beq
  4 \pi^2 f_X^2 &\simeq& M^2 \, e^{m_X^2/M^2} \left(1-  e^{-s_0/M^2}\right) \,,
\eeq
where $M^2 \equiv \omega_M \, (n_+p')$ and $s_0 \equiv \omega_s \, (n_+p')$,
we can write the above result as
\beq
  \Delta F_X(\mu,n_+p') &=&  \frac{8 \pi^2 f_B f_X}{m_B}
   \int_0^\infty \frac{d\omega}{\omega} \, \phi_B^+(\omega,\mu)
   \frac{1- e^{-\frac{s_0}{M^2}} \, \theta(\omega-\frac{s_0}{n_+p'})
          - e^{-\omega n_+p'/M^2} \, \theta(\frac{s_0}{n_+p'}-\omega)}
   {1-  e^{-\frac{s_0}{M^2}}} \,.
\cr && \label{eq:DeltaF2} \eeq
Notice that, in contrast to the
case for the non-factorizable form factor $\xi_M$, the limit $s_0
\ll \omega \, n_+p'$ now does exist. This implies that the
soft dynamics related to the $B$\/-meson constituents decouples from
the dynamics related to the spectrum of the interpolating
current. Therefore, it is indeed justified to identify the
sum rule parameters in the otherwise independent sum rules
for $\Delta F_X$ and the corresponding decay constant $f_X$.
Not surprisingly, in this limit the above
formula (\ref{eq:DeltaF2}) reduces to the prediction
from QCD factorization, with -- at this order -- the
asymptotic form for the light meson LCDAs $\phi_X(u)$,
\beq
  \Delta F_X(\mu) \Big|_{\rm QCDF, asympt.} &=&
   \frac{8\pi^2 f_B f_X}{m_B}
   \int_0^\infty \frac{d\omega}{\omega} \, \phi_B^+(\omega,\mu) \,.
\label{eq:DeltaF3}
\eeq

%%%%%%%%%%%%%%%%%%%%%%%%%%%%%%%%%%%%%%%%%%%%%%%%%%%%%%%%%%%%%%%%%%%%%%%%5
%%%%%%%%%%%%%%%%%%%%%%%%%%%%%%%%%%%%%%%%%%%%%%%%%%%%%%%%%%%%%%%%%%%%%%%%%

\section{Numerical Predictions for form factors}\label{numanalysis}

%%%%%%%%%%%%%%%%%%%%%%%%%%%%%%%%%%%%%%%%%%%%%%%%%%%%%%%%%%%%%%%%%%%%%%%%%%
%%%%%%%%%%%%%%%%%%%%%%%%%%%%%%%%%%%%%%%%%%%%%%%%%%%%%%%%%%%%%%%%%%%%%%%%%

\subsection{Hadronic input parameters}

%%%%%%%%%%%%%%%%%%%%%%%%%%%%%%%%%%%%%%%%%%%%%%%%%%%%%%%%%%%%%%%%%%%%%%%%%%

The result for the soft form factors from the SCET sum rules
depends on the following hadronic input parameters:
\begin{itemize}
 \item The $B$\/-meson distribution amplitude
       $f_B \, \phi_B^-(\omega,\mu)$: In the following,
       we will relate $1/\omega_0 =\phi_B^-(0)$ to
       $\lambda_B=\displaystyle\Big[ \int_0^\infty d \omega \, {\phi_+ (\omega) \over \omega }
       \Big]^{-1}$ via the
       Wandzura-Wilczek approximation \cite{Beneke:2000wa}
       at the low scale $\mu = 1$~GeV.
       We take the values which have been obtained
       from a recent moment analysis in \cite{Lee:2005gz}
$$
  \omega_0(1~{\rm GeV}) = (0.48 \pm 0.05)~{\rm GeV} \,.
$$
(The value is consistent with the sum-rule result from
 \cite{Braun:2003wx}.) For the shape of $\phi_B^-(\omega)$,
we adopt the simple parametrization
\beq
 \phi_B^-(\omega) &=& \frac{1}{\omega_0} \, e^{-\omega/\omega_0} \,,
\label{phidef}
\eeq
see, however, also the discussion in Section~\ref{modelphi}.

\item The $B$\/-meson decay constant is taken as
      $f_B(m_b) = (180 \pm 30)$~MeV
      which corresponds to $f_B(1~{\rm GeV})
      \simeq (150 \pm 30)$~MeV.

\item The threshold parameter $\omega_s = s_0/(n_+p')$:
      The next vector-meson resonance above the $\rho$
      meson is the $\rho(1450)$, and therefore we may expect
      $\omega_s \sim 0.4$~GeV at $(n_+p')=m_B$.
      We will take
      $\omega_s=\{0.35,0.4,0.45\}$~GeV
      as our default range
      for longitudinal $\rho$-mesons.\footnote{In the following,
      all values assigned to the sum-rule parameters $\omega_s$
      and $\omega_M$ refer to the maximal recoil point $(n_+p')=m_b$.
      However, when $(n_+p')$ or $m_Q \neq m_b$ are varied, these
      parameters are re-scaled accordingly.}

      In case of the correlation function $\Pi_\perp$ for transversely
      polarized $\rho$ mesons, also the axial-vector resonances
      may contribute, starting with $b_1(1235)$.
      As a consequence, the threshold parameter in our sum rule
      for $\xi_\perp$ is expected to be significantly smaller
      than in the longitudinal case
      (see, for instance, the discussion in \cite{Ball:1996tb}).
      Therefore, we consider the lower threshold values,
      $\omega_s=\{0.20,0.25,0.30\}$~GeV for transverse $\rho$-mesons.

For completeness, we will also reconsider the soft $B\to \pi$ form factor,
for which we will take $\omega_s = \{0.15,0.20,0.25\}$~GeV.

\item The Borel parameter $\omega_M$: Reasonable values of $\omega_M$
      should be estimated from the sum rule itself. As it turns out, the
      prediction for the soft form factors approximately grows logarithmically
      with $\omega_M$.  Requiring that the sum rule is sufficiently stable against
      variations of $\omega_M$ thus determines a {\em lower}\/ acceptable value
      for $\omega_M$. In practice, we consider the normalized logarithmic
      derivative
      \beq
        D &=& \frac{\omega_M}{\xi_{\parallel,\perp,\pi}} \,
         \frac{\partial \, \xi_{\parallel,\perp,\pi}}{\partial \omega_M}
      \eeq
      and require $|D| < 25\%$ (however, we generally do not consider
       values of $\omega_M$ below $m_\rho^2/m_b \simeq 0.1$~GeV).
      As a further constraint, we impose that the continuum contribution
     ($\omega' > \omega_s$) is  not too large, and require that
      \beq
      R &=& \frac{\int\limits^{\omega_s}_0 d\omega' \,
      e^{-\omega'/\omega_M} \,
      \phi^{\rm eff}(\omega')}
      {\int\limits_{0}^\infty d\omega' \,  e^{-\omega'/\omega_M} \,
      \phi^{\rm eff}(\omega')}
      \eeq
      is larger than 50\%.

\item For the meson decay constants\footnote{In the following, we do
not quote the uncertainty w.r.t.\ the decay constant $f_\rho^\perp$
which amounts to about 5-10\% and would propagate linearly
into the form factor sum rule.} we take $f_\rho^\parallel = 205$~MeV,
      and $f_\rho^\perp(\mu=1~{\rm GeV}) = 160 $~MeV
      \cite{Ball:1996tb}.

\end{itemize}

\subsection{Soft form factor for longitudinal vector mesons}

\begin{figure}[t!!]
\begin{center}
(a)\hspace{-2em}
\psfrag{x}{\hspace{-1.7em}\footnotesize \parbox{5em}{\vspace{0.5em}$\omega_M$ (GeV)} }
\psfrag{y}{\hspace{-0.7em} \footnotesize \parbox{2em}{$\hat\xi_\parallel$} }
\includegraphics[height=0.27\textwidth]{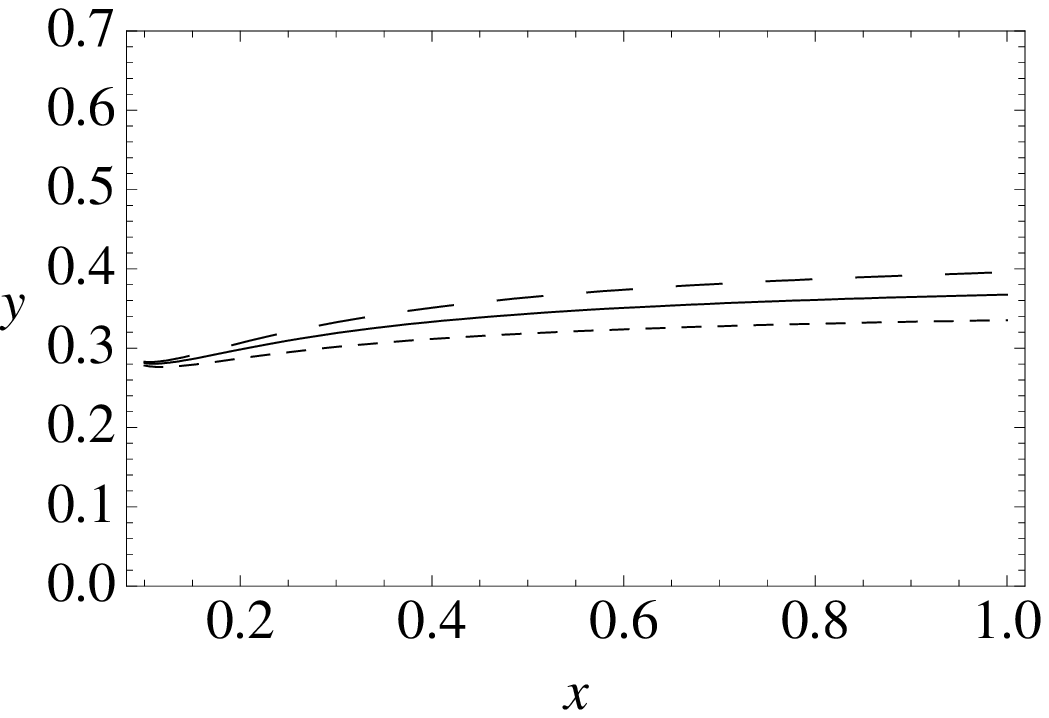} \hspace{2em}
\hspace{2em}
(b)\hspace{-2em}
\includegraphics[height=0.27\textwidth]{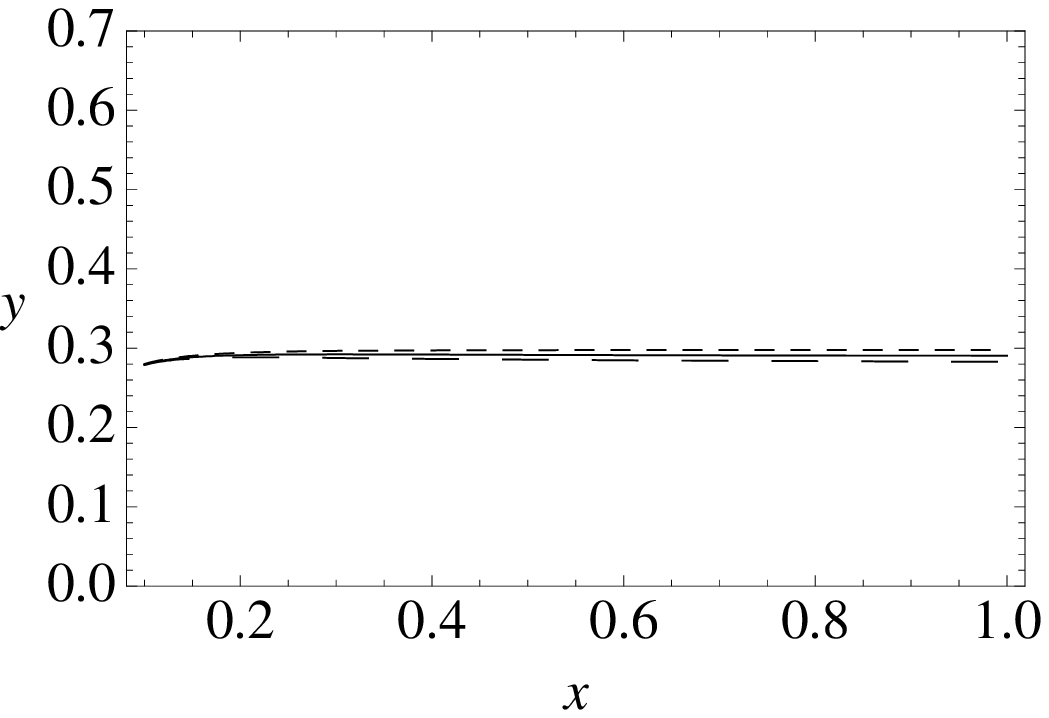}
\end{center}
\centerline{\parbox{0.95\textwidth}{\caption{\small (a) The soft form factor $\hat \xi_\parallel$ at $n_+p'=m_B$
and $\mu=1.0$~GeV as a function of
the Borel parameter for three different values of the threshold
parameters (solid line: $\omega_s=0.4$~GeV, short-dashed line:
$\omega_s=0.45$~GeV, long-dashed line: $\omega_s=0.35$~GeV). (b) The
same quantity using the modified sum rule (\ref{eq:phieff_neu}).}
\label{fig:par_Borel}}}
\end{figure}

Fig.~\ref{fig:par_Borel}(a) shows the dependence of the soft form
factor $\hat \xi_\parallel$ at $n_+p'=m_B$
and $\mu=1.0$~GeV as a function of the Borel parameter for three
different values of the threshold parameters.
Our constraints $R_\parallel>50\%$ and $D<25\%$ are fulfilled for
a rather large range of Borel parameter values,
$  0.1~{\rm GeV} < \omega_M < 3.35~{\rm GeV} \,,
$
and we take the geometric mean
$\omega_M = 0.6~{\rm GeV}$ as our central value.
Taking into account the variation of the various input parameters, we obtain
the estimate
\beq
\hat\xi_\parallel(n_+p'=m_B,\mu=1.0~{\rm GeV})
 &=&
  0.33 {}^{+0.02}_{-0.02} \big|_{\omega_s}
     {}^{+0.03}_{-0.06} \big|_{\omega_M}
     {}^{+0.03}_{-0.02} \big|_{\omega_0}
     {} \pm 0.05 \big|_{f_B}\,.
\eeq Adding errors in quadrature amounts to $\hat
\xi_\parallel(n_+p'=m_B,\mu=1.0~{\rm GeV}) =
0.33^{+0.07}_{-0.09}$, i.e.\ in total a $\sim25\%$ uncertainty
with the largest contribution coming from the Borel parameter and
the decay constant. For
comparison, the tree level result for the same input parameters
would give $\hat\xi_\parallel=0.38$, illustrating the
numerical significance of the $\alpha_s$ corrections.
Within the rather large uncertainties, our
value is compatible with the values obtained from traditional sum
rules \cite{Ball:1998kk,Ball:2004rg}. (For instance, the phenomenological
analysis in \cite{Beneke:2004dp} (see Table~2 therein) infers
$\hat\xi_\parallel(m_B,1.5~{\rm GeV}) = 0.37 \pm 0.06$.)

In principle, one can decrease the
sensitivity on the sum rule parameters by dividing the sum rule for
$\xi_\parallel$ by the leading-order sum rule for the decay
constant $f_\rho$,
$$
 4 \pi^2 \, f_\rho^2 \, e^{-m_\rho^2/M^2} \simeq
 \int_0^{s_0} ds \, e^{-s/M^2} \,,
$$
where, for simplicity, we have not included $\alpha_s$ corrections
and non-perturbative corrections from quark and gluon condenstates,
which are suppressed by powers of $1/M^2$.
If we {\em assume}\/ that the parameters $\omega_M = M^2/(n_+p')$ and
$\omega_s=s_0/(n_+p')$ in the two sum rules can be identified, we
obtain a modified sum rule
\beq
  \hat \xi_\parallel(n_+p',\mu) &\simeq &
  \frac{4 \pi^2 f_\rho f_B m_B \,
   \int\limits_0^{\omega_s} d\omega' \, e^{-\omega'/\omega_M} \,
   \phi_\parallel^{\rm eff}(\omega',n_+p',\mu)}
  { (n_+ p')^2 \, \int\limits_0^{\omega_s} d\omega' \, e^{-\omega'/\omega_M}
}\,. \label{eq:phieff_neu} \eeq Numerically, the modified sum rule
would result in the estimate \beq \hat
\xi_\parallel(n_+p'=m_B,\mu=1.0~{\rm GeV})
 &=&
  0.29 {}\pm 0.01 \big|_{\omega_s}
     {}^{+0.00}_{-0.01} \big|_{\omega_M}
     {}\pm 0.02 \big|_{\omega_0}
     {}\pm 0.05 \big|_{f_B}
     {}\pm (?)_{\rm syst.}\,.
\nonumber
\eeq
The result is plotted in
Fig.~\ref{fig:par_Borel}(b). It is to be stressed, however, that
the very fact that the soft and collinear dynamics in
$\xi_\parallel$ do not factorize in QCD also implies that the sum
rule parameters for the $B \to \rho$ form factor and the
$\rho$\/-meson decay constant are not trivially correlated,
and the above numerical value is only shown for illustration.
The reduction of the $\omega_M$ dependence in
(\ref{eq:phieff_neu}) is probably compensated by an increased
systematic error (indicated by the question mark)
which is hard to estimate reliably.
Nevertheless, we will find the modified sum rules useful in
the context of form factor ratios (see section~\ref{sec:ffratio}),
where one could argue that the systematic error from correlating
parameters in different sum rules and neglecting higher-order
corrections drops out to some extent.

\subsubsection{Energy dependence}

\begin{figure}[tph]
\begin{center}
\psfrag{x}{\hspace{-1.7em}\footnotesize \parbox{6.5em}{\vspace{0.5em}$(n_+p')$ (GeV)} }
\psfrag{y}{\hspace{-2.9em}
 \parbox{2em}{$\frac{\hat\xi_\parallel(n_+p')}{\hat\xi_\parallel(m_B)}$} }
(a)
\includegraphics[height=0.27\textwidth]{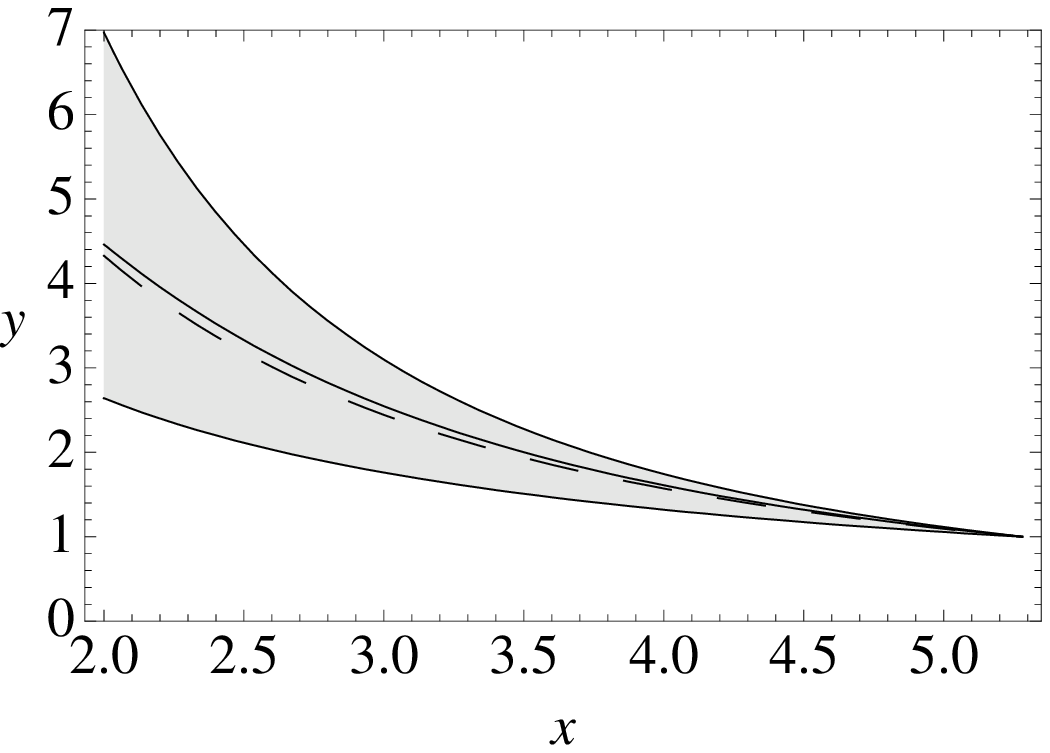}
\hspace{4em}
(b)
\psfrag{x}{\hspace{-2.2em}\footnotesize \parbox{7.5em}{\vspace{0.5em}$1/m_Q$ (GeV$^{-1}$)} }
\psfrag{y}{\hspace{-2.9em} \footnotesize \parbox{2em}{$R_\parallel(m_Q)$} }
\hspace{-2em}
\includegraphics[height=0.27\textwidth]{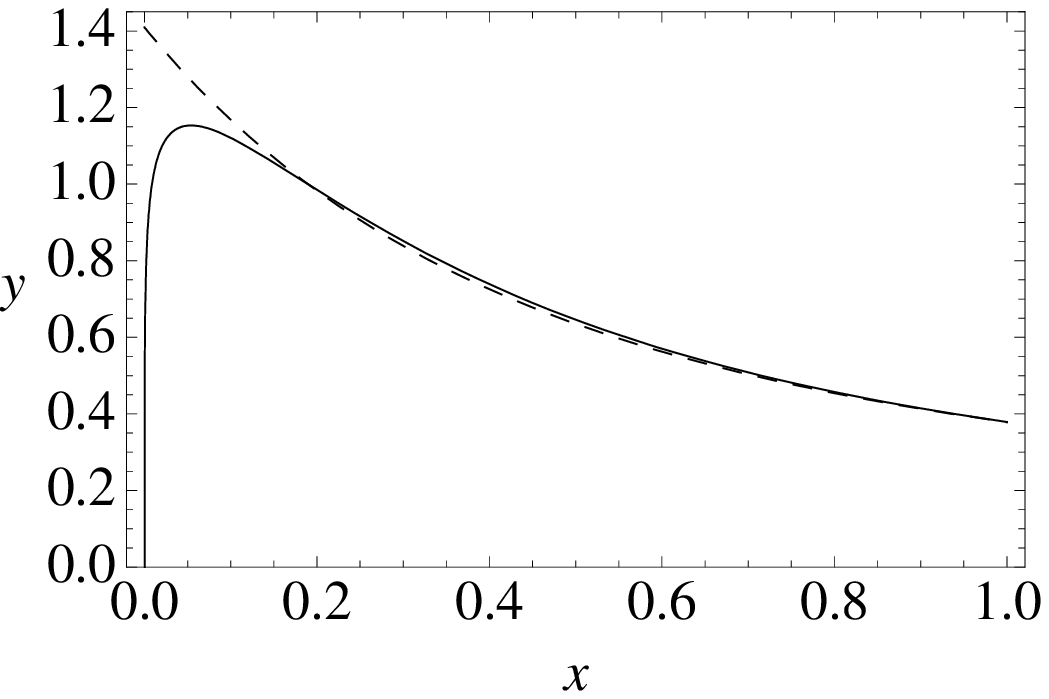}
\end{center}
\centerline{\parbox{0.95\textwidth}{\caption{\small (a) Energy dependence of the soft form factor
$\hat\xi_\parallel(n_+p')/\hat\xi_\parallel(m_B)$.
The grey band illustrates the range between a pure $1/(n_+p')^2$
and a pure $1/(n_+p')$ behaviour.
(b)
The $m_Q$ dependence of the soft form factor $R_\parallel(m_Q)$, see (\ref{eq:Rpar}),
as a function of $1/m_Q$. The solid line denotes
the NLO result, the dashed line the LO result.}
\label{fig:mbdep_par}
\label{fig:endeppar}}}
\end{figure}

We may also study the energy dependence of the soft form
factor, which is of phenomenological importance if one wants
to interpolate between lattice-theory results at intermediate
values of $q^2$ and sum-rule results for the large-recoil
limit at $q^2=0$ (for a comprehensive discussion, see
e.g.\ \cite{Hill:2005ju}).
The normalized result for the energy dependence is shown in
Fig.~\ref{fig:endeppar}(a), where we have also shown a pure $1/(n_+p')$
and $1/(n_+p')^2$ behaviour for comparison. Here we considered the sum
rule parameters to scale with the recoil energy according to
$M^2 = \omega_M \, (n_+p')={\rm const.}$ and
$s_0 = \omega_s \, (n_+p')={\rm const.}$
We observe that the resulting energy dependence is just between a $1/(n_+p')$
and a $1/(n_+p')^2$ fall off. We also note that the difference between the
energy dependence of the
tree-level and the NLO result is tiny. We have also studied the stability
criteria of the sum rule as a function of $(n_+p')$ and found that the
$\omega_M$ dependence is not changed very much, while the
continuum pollution is even decreasing with smaller values of $(n_+p')$.
We therefore consider the predictions for the
energy dependence of the soft form factor $\hat \xi_\parallel$ to be
rather solid.

%%%%%%%%%%%%%%%%%%%%%%%%%%%%%%%%%%%%%%%%%%%%%%%%%%%%%%%%%%%%%%%

\subsubsection{$m_Q$ dependence}

\label{sec:mQ}

The scaling of the soft form factor with the heavy-quark mass
is another interesting issue. From the theoretical side one
is interested in the competition between the soft Feynman
mechanism for intermediate values of $m_Q$ and the Sudakov
supression in the asymptotic limit $m_Q \to \infty$.
Phenomenologically, the heavy-quark mass scaling could be exploited to
get a rough estimate for $D$\/-meson decays at large recoil by extrapolation from
the $B$\/-meson case.
In Fig.~\ref{fig:endeppar}(b) we study the $m_Q$ dependence of
the normalized soft form factor at maximal recoil $(n_+p')=m_Q$,
\beq
  R_\parallel(m_Q) &\equiv& \frac{m_Q}{m_B} \, \frac{f_B(\mu)}{f_Q(\mu)} \,
  \frac{\hat \xi_\parallel(E_{\rm max})|_{m_Q}}
       {\hat \xi_\parallel(E_{\rm max})|_{m_B}} \,,
\label{eq:Rpar}
\eeq
where the factorization scale is kept fixed at $\mu=1.0$~GeV,
and the pre-factors take into account the ``trivial'' $m_Q$ dependence.
Again, the sum rule parameters are scaled such that $M^2 = \omega_M \, m_Q={\rm const.}$
and $s_0 = \omega_s \, m_Q = {\rm const.}$

For values $m_Q < m_b$, the situation is analogous to the energy dependence ($n_+p' < m_B$)
discussed above, i.e.\ an extrapolation of the sum rule to smaller heavy quark masses
seems to be justified. However, the $1/m_Q$ power corrections become,
of course, even more important in that case (see also the discussion in
\cite{Khodjamirian:2005ea}). On the other hand, for larger values
of $m_Q$, the sum rule predictions becomes more and more unstable. This manifests
itself in: (i) a huge difference between the tree-level and the NLO result in
the limit $m_Q \to \infty$, (ii) an increased sensitivity to the sum rule parameters,
(iii) an increased continuum pollution.
In particular, the very heavy quark limit $m_Q \to \infty$ for $\hat \xi_\parallel$
does not exist anymore beyond the tree-level approximation, which can be traced back
to the non-factorizable large logarithms discussed in Sec.~\ref{largelog}. However,
one has to keep in mind that only the product of the
soft form factor $\hat \xi_\parallel(n_+p',\mu)$ and the matching coefficient
$C_i^{\rm I}(\mu,n_+p')$
between the QCD and the SCET current contributes to the physical form factor
in (\ref{factorization}), and $C_i^{\rm I}(\mu,m_Q)$
vanishes in the limit $m_Q \to \infty$ (for fixed values of $\mu$)
due to the exponentiation of Sudakov logs \cite{Bauer:2000yr}.

%%%%%%%%%%%%%%%%%%%%%%%%%%%%%%%%%%%%%%%%%%%%%%%%%%%%%%%%%%%%%%%%%%%%%%%%%%%%

\subsection{Soft form factor for transverse vector mesons}

%%%%%%%%%%%%%%%%%%%%%%%%%%%%%%%%%%%%%%%%%%%%%%%%%%%%%%%%%%%%%%%%%%%%%%%%%%%%

\begin{figure}
\begin{center}
(a)\hspace{-2em}
\psfrag{x}{\hspace{-1.7em}\footnotesize \parbox{5em}{\vspace{0.5em}$\omega_M$ (GeV)} }
\psfrag{y}{\hspace{-0.8em} \footnotesize \parbox{2em}{$\xi_\perp$} }
\includegraphics[height=0.27\textwidth]{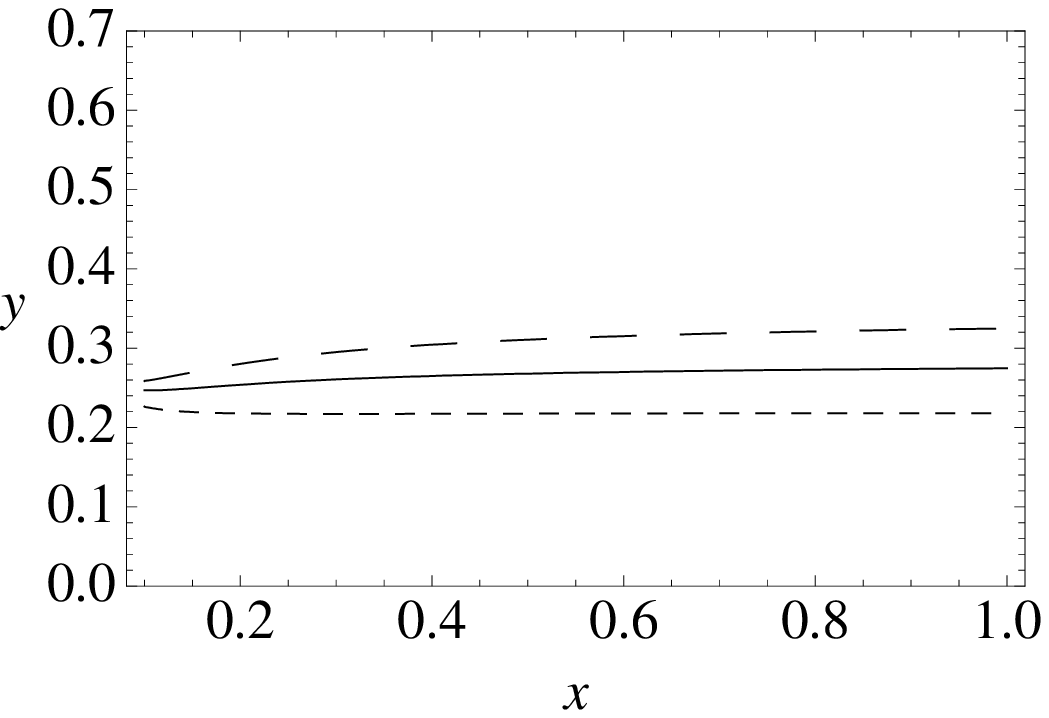} \hspace{2em}
\hspace{2em}
(b)\hspace{-2em}
\includegraphics[height=0.27\textwidth]{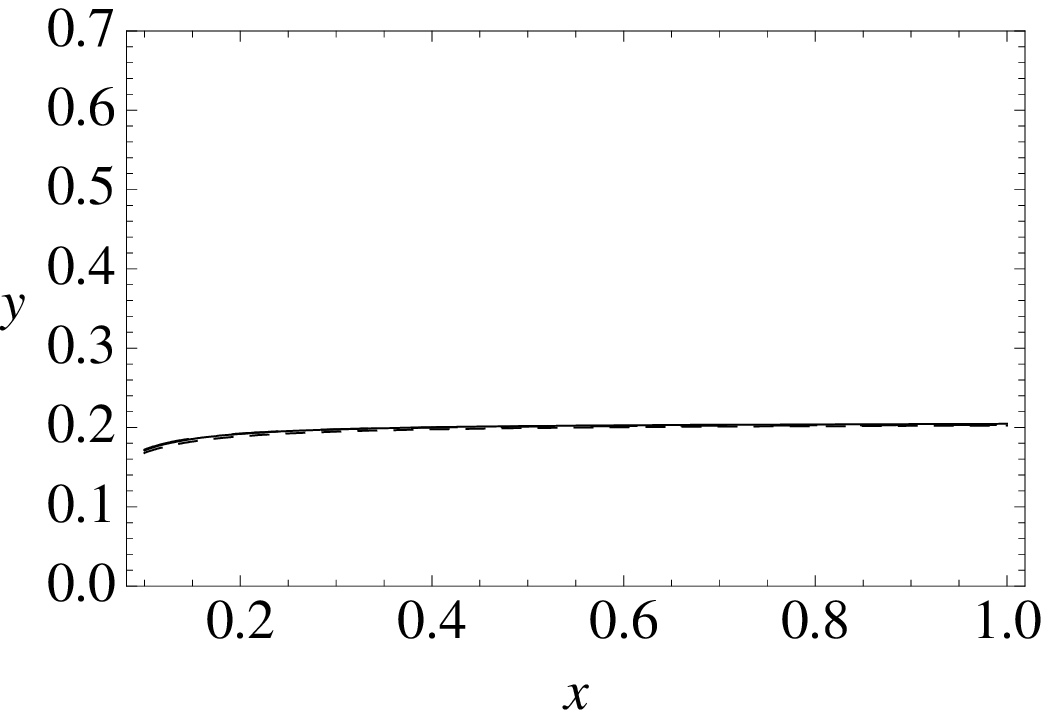}
\end{center}
\centerline{\parbox{0.95\textwidth}{\caption{\small (a) The soft form factor $\xi_\perp$ at $n_+p'=m_B$ and
$\mu=1.0$~GeV as a function of the Borel parameter for three
different values of the threshold parameters (solid line:
$\omega_s=0.25$~GeV, short-dashed line: $\omega_s=0.2$~GeV,
long-dashed line: $\omega_s=0.3$~GeV). (b) The same quantity
using the modified sum rule (\ref{eq:phieffp_neu}).}
\label{fig:perp_Borel}}}
\end{figure}

The analysis of the form factors for $B$\/-decays into
transversely polarized vector mesons is similar as for
longitudinal ones. The main difference is the smaller value
for the default threshold parameter $\omega_s$, due to the
contribution of the $b_1(1235)$ to the correlator $\Pi_\perp$.
The numerical result for the
form factor $\xi_\perp$  is
shown in Fig.~\ref{fig:perp_Borel}(a). Considering the stability of
the sum-rule result for $\xi_\perp$ and the smallness of the
continuum contribution, we find that also the Borel parameter
$\omega_M$ has to be chosen in a range smaller than in the
longitudinal case, $0.1 < \omega_M < 0.8$~GeV, with
the central value taken as $\omega_M =0.3$~GeV.
Our result for the soft transverse $B \to \rho$ form factor
follows as
\beq
  \xi_\perp(n_+p'=m_B,\mu=1.0~{\rm GeV})
 &=&
  0.26 {}^{+0.03}_{-0.04} \big|_{\omega_s}
     {}^{+0.01}_{-0.01} \big|_{\omega_M}
     {}\pm 0.03 \big|_{\omega_0}
     {}\pm 0.04 \big|_{f_B}
\eeq
Adding errors in quadrature amounts to
$\xi_\perp(n_+p'=m_B,\mu=1.0~{\rm GeV}) = 0.26^{+0.06}_{-0.07}$, i.e.\ in
total a $\sim 25\%$ uncertainty with similar contributions
from the different input parameters. The tree-level
result for the same input parameters gives $\xi_\perp=0.39$.
(For comparison, in \cite{Beneke:2004dp}
 the value $\xi_\perp(\mu=1.5~{\rm GeV}) = 0.27 \pm 0.05$
 has been derived from the traditional sum rule approach in \cite{Ball:1998kk}.)

\begin{figure}[t!!bp]
\begin{center}
\psfrag{x}{\hspace{-1.7em}\footnotesize \parbox{6.5em}{\vspace{0.5em}$(n_+p')$ (GeV)} }
\psfrag{y}{\hspace{-3.0em}
 \parbox{2em}{$\frac{\xi_\perp(n_+p')}{\xi_\perp(m_B)}$} }
(a)
\includegraphics[height=0.27\textwidth]{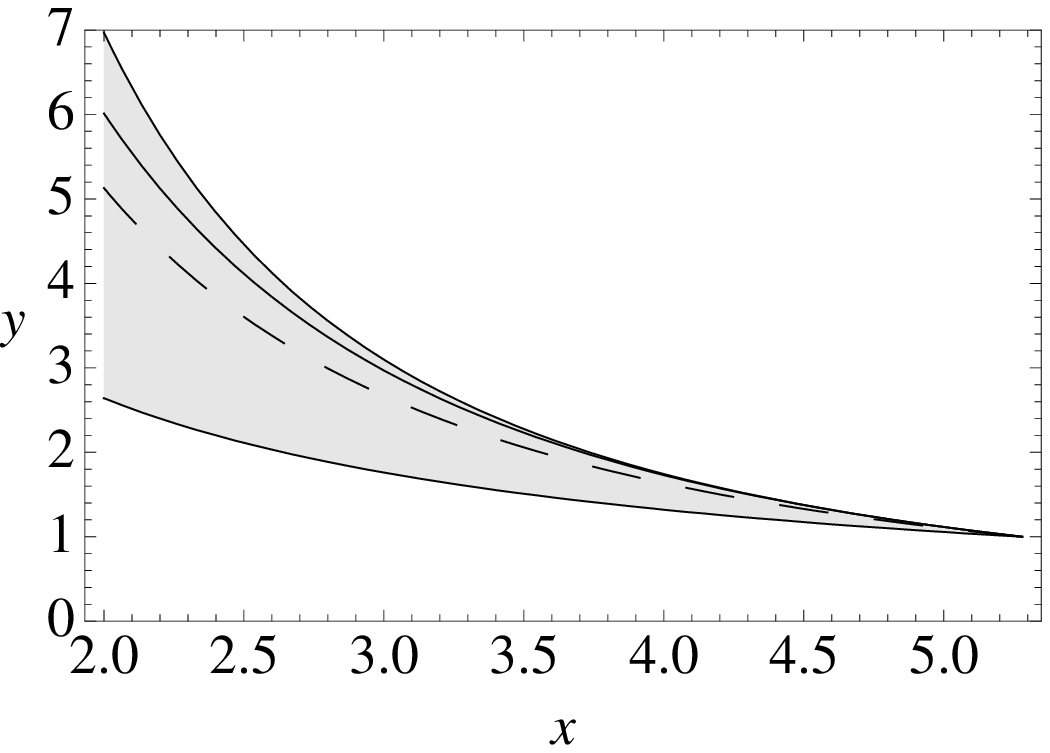}
\hspace{4em}
(b)
\psfrag{x}{\hspace{-2.2em}\footnotesize \parbox{7.5em}{\vspace{0.5em}$1/m_Q$ (GeV$^{-1}$)} }
\psfrag{y}{\hspace{-3.0em} \footnotesize \parbox{2em}{$R_\perp(m_Q)$} }
\hspace{-2em}
\includegraphics[height=0.27\textwidth]{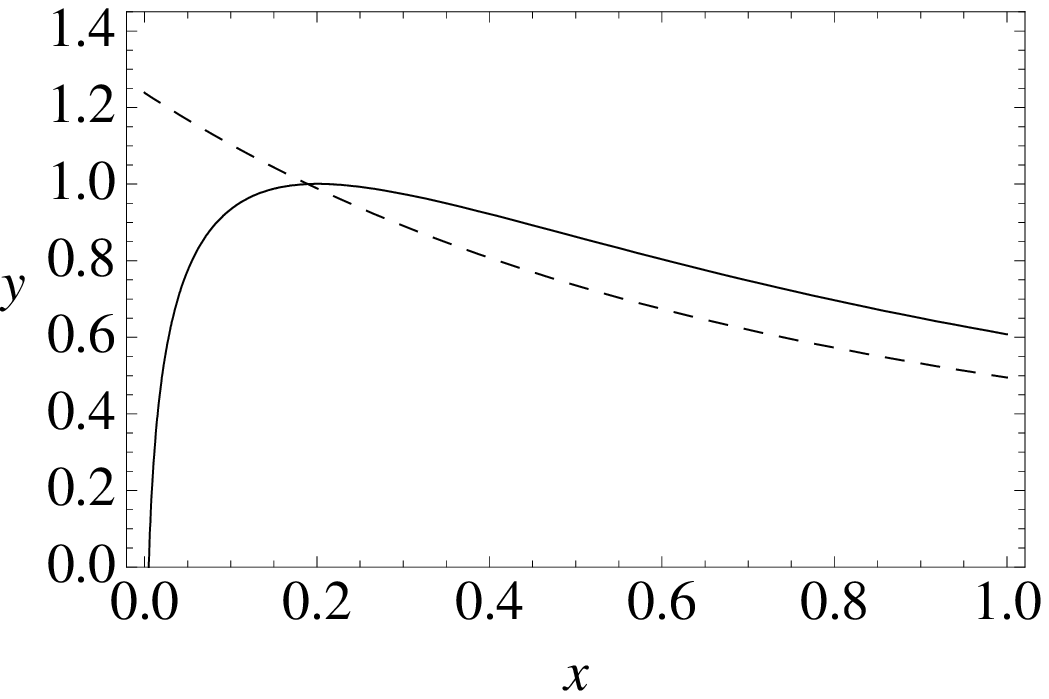}
\end{center}
\centerline{\parbox{0.95\textwidth}{\caption{\small (a) Energy dependence of the soft form factor
$\xi_\perp(n_+p')/\xi_\perp(m_B)$.
The grey band illustrates the range between a pure $1/(n_+p')^2$
and a pure $1/(n_+p')$ behaviour.
(b)
The $m_Q$ dependence of the soft form factor $R_\perp(m_Q)$,
(defined analogously as (\ref{eq:Rpar})),
as a function of $1/m_Q$. The solid line denotes
the NLO result, the dashed line the LO result.}
\label{fig:endepperp}}}
\end{figure}

If we divide the sum rule for $\xi_\perp$ by the corresponding
sum rule for $f_{\rho}^\perp$ \cite{Ball:1996tb},
$$
  4 \pi^2 \,  (f_\rho^\perp) ^2 \, e^{-m_\rho^2/M^2} \simeq
 \int_0^{s_0} ds \, e^{-s/M^2}
 \left(1 + \frac{\alpha_s}{\pi} \left[\frac79 + \frac23 \ln \frac{s}{\mu^2} \right] \right) \,,
$$
we obtain the modified prediction
\beq
  \xi_\perp(n_+p',\mu) &= &
  \frac{4 \pi^2 f_\rho^\perp f_B m_B \,
   \int\limits_0^{\omega_s} d\omega' \, e^{-\omega'/\omega_M} \,
   \phi_\perp^{\rm eff}(\omega',n_+p',\mu)} { (n_+ p')^2 \, \int\limits_0^{\omega_s} d\omega' \, e^{-\omega'/\omega_M}
  \left( 1 + \frac{\alpha_s}{\pi} \left(\frac79 + \frac23 \ln \frac{\omega'
        n_+p'}{\mu^2} \right)\right)} \,,
\label{eq:phieffp_neu} \eeq where we also took into account the
explicit $\mu$ dependence arising from the ${\cal O}(\alpha_s)$
correction to the tensor current. Again, the modified sum rule
reduces the parametric uncertainties related to $\omega_s$ and
$\omega_M$, but induces an unknown systematic error due to the
assumed correlation between the two sum rules,
\beq
\xi_\perp(n_+p'=m_B,\mu=1.0~{\rm GeV})
 &=&
  0.20 {}^{+0.00}_{-0.00} \big|_{\omega_s}
     {}^{+0.01}_{-0.03} \big|_{\omega_M}
     {}\pm 0.02 \big|_{\omega_0}
     {}\pm 0.03 \big|_{f_B}
     {}\pm (?)_{\rm syst.}
\nonumber \eeq Notice that the central value obtained from the
modified sum rule is significantly smaller than the one from the
original sum rule (and only marginally consistent within the
uncertainties). 
In Fig.~\ref{fig:endepperp} we
show again the energy and heavy-mass dependence.
%, which is qualitatively similar to the longitudinal case.
Compared to the longitudinal case, the energy dependence of $\xi_\perp$ is somewhat closer to a $1/(n_+p')^2$ fall-off.
Also the heavy-quark mass dependence is slightly different.

%%%%%%%%%%%%%%%%%%%%%%%%%%%%%%%%%%%%%%%%%%%%%%%%%%%%%%%%

\subsection{Soft form factor for pseudoscalar mesons}

%%%%%%%%%%%%%%%%%%%%%%%%%%%%%%%%%%%%%%%%%%%%%%%%%%%%%%%%

For completeness, we also show the results for the soft
$B\to \pi$ form factor, where the sum rule takes the same
form as for longitudinally polarized vector mesons
(with $m_\pi^2 \approx 0$),
\beq
  \xi_\pi(n_+p',\mu) &= &
  \frac{f_B m_B}{f_\pi \, (n_+p')} \,
   \int\limits_0^{\omega_s} d\omega' \, e^{-\omega'/\omega_M} \,
   \phi_\parallel^{\rm eff}(\omega',n_+p',\mu)\,.
\label{eq:phipi} \eeq
The choice for the threshold parameter $s_0$ is somewhat more difficult
in this case, since the axial-vector current, used to interpolate the pion,
is expected to receive sizeable contributions from multi-pion channels.
Typical values for $s_0$ used in sum rules for, say,
the pion decay constant are thus rather low and lie in the range $0.7 -1.0$~GeV$^2$,
corresponding to $\omega_s = 0.15 - 0.20$~GeV. On the other hand, one might
argue that the Feynman mechanism for the soft $B \to X$ form factor is dominant
for single-particle states $X$, while multi-particle decays $B \to \pi\pi\pi$ etc.\
should be suppressed. In this case, one would expect values of $s_0$ in the
$B \to \pi$ sum rule close to the first axial-vector resonance.
In the following, we choose a rather conservative range
$\omega_s = \{0.15,0.2,0.25\}$~GeV (in accordance with our previous
work \cite{DeFazio:2005dx}). Consequently, our result for $\xi_\pi$
will have a rather large uncertainty from the variation of $s_0$.

As for the Borel parameter, we again consider the
stability of $\xi_\pi$ with respect to variations of $\omega_M$
and the amount of continuum pollution within the given range for $\omega_s$.
From $D < 25\%$ and $R > 50\%$ we infer $0.3~{\rm GeV} < \omega_M < 0.7~{\rm GeV}$
with the central value chosen as $\omega_M =0.45$~GeV. Notice that the allowed range
for $\omega_M$ is somewhat smaller than in the vector meson case, which is due to
the rather small values of $s_0$. With even smaller values of $s_0$ (like e.g.\ the ones
chosen in \cite{Chay:2005gh}), the criteria $D < 25\%$ and $R > 50\%$ could not be
simultaneously fulfilled anymore. Moreover, for too small values of $\omega_s$ the
convergence of the perturbative series for the sum rule is bad, since the $\alpha_s$
corrections are enhanced by large logarithms $\ln \omega_s/\omega_0$ (see the discussion
above). This again could be taken as an indication that the considered lower values for
$s_0$ are less realistic.

\begin{figure}[t!!]
\begin{center}
(a)\hspace{-2em}
\psfrag{x}{\hspace{-1.7em}\footnotesize \parbox{5em}{\vspace{0.5em}$\omega_M$ (GeV)} }
\psfrag{y}{\hspace{-0.8em} \footnotesize \parbox{2em}{$\xi_\pi$} }
\includegraphics[height=0.27\textwidth]{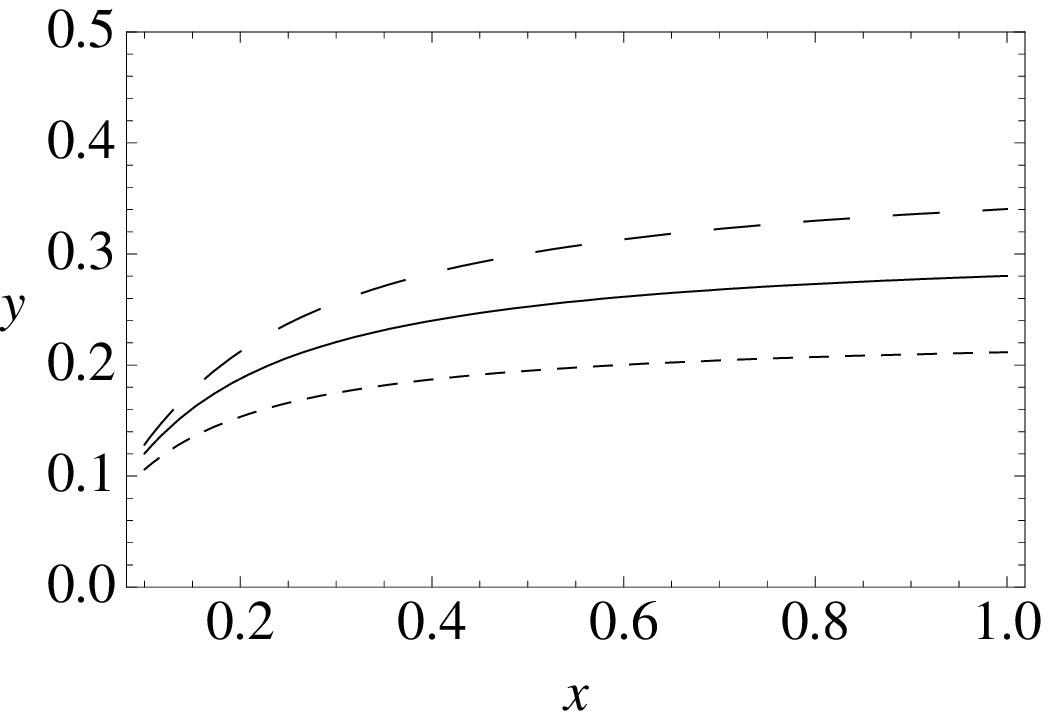} \hspace{2em}
\hspace{2em}
(b)\hspace{-2em}
\includegraphics[height=0.27\textwidth]{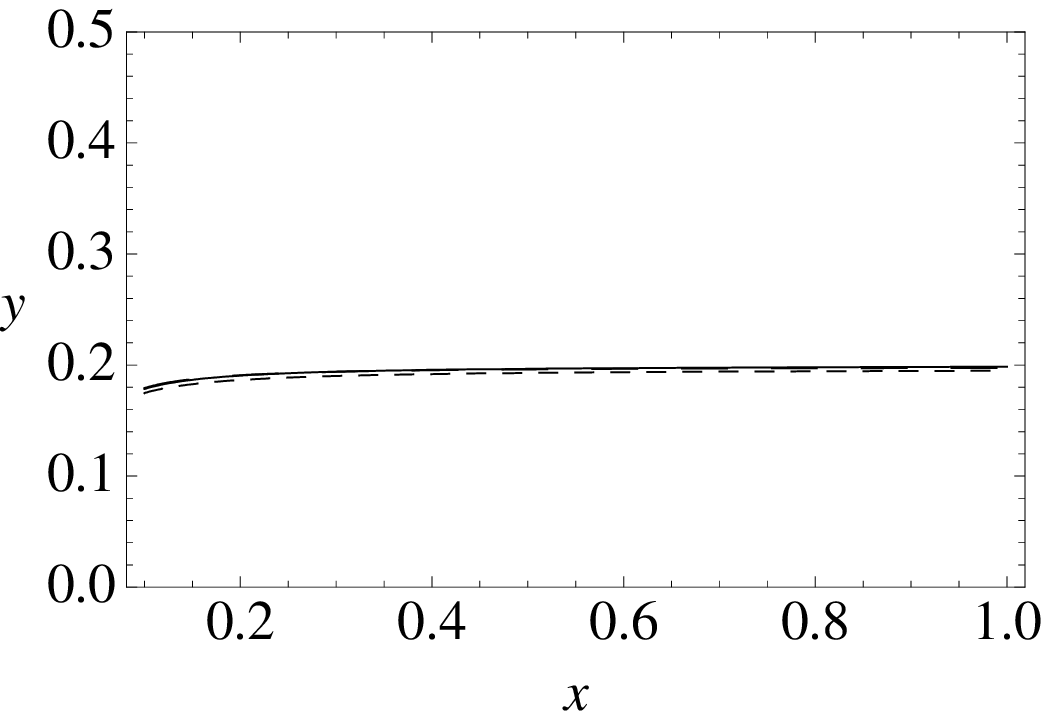}
\end{center}
\centerline{\parbox{0.95\textwidth}{\caption{\small (a) The soft form factor $\xi_\pi$ at $n_+p'=m_B$ and
$\mu=1.0$~GeV as a function of the Borel parameter for three
different values of the threshold parameters (solid line:
$\omega_s=0.2$~GeV, short-dashed line: $\omega_s=0.25$~GeV,
long-dashed line: $\omega_s=0.15$~GeV). (b) The same quantity
using the modified sum rule (\ref{eq:phieffpi_neu}).}
\label{fig:pi_Borel}}}
\end{figure}

In any case, our choice for the variation of the sum-rule parameters
implies the following estimate for $\xi_\pi$ at maximal recoil,
\beq
  \xi_\pi(n_+p'=m_B,\mu=1.0~{\rm GeV})
 &=&
  0.25 {}^{+0.05}_{-0.06} \big|_{\omega_s}
     {}^{+0.02}_{-0.03} \big|_{\omega_M}
     {}^{+0.03}_{-0.02} \big|_{\omega_0}
     {}\pm 0.04 \big|_{f_B}\,.
\eeq Adding errors in quadrature amounts to
$\xi_\pi(n_+p'=m_B,\mu=1.0~{\rm GeV}) = 0.25^{+0.07}_{-0.08}$,
i.e.\ in total a $\sim 30\%$ uncertainty with the largest
contributions coming from the decay constant and the threshold
parameter. (Again, we quote the corresponding tree-level result,
$\xi_\pi=0.32$, for comparison).

The modified sum rule is obtained by dividing the
sum rule for $f_\pi$,
\beq
    \xi_\pi(n_+p',\mu) &\simeq &
  \frac{4 \pi^2 f_\pi f_B m_B \,
   \int\limits_0^{\omega_s} d\omega' \, e^{-\omega'/\omega_M} \,
   \phi_\parallel^{\rm eff}(\omega',n_+p',\mu)}
  { (n_+ p')^2 \, \int\limits_0^{\omega_s} d\omega' \, e^{-\omega'/\omega_M}
}\,. \label{eq:phieffpi_neu} \eeq Numerically, we obtain \beq
  \xi_\pi(n_+p'=m_B,\mu=1.0~{\rm GeV})
 &=&
  0.20 {}^{+0.00}_{-0.00} \big|_{\omega_s}
     {}^{+0.00}_{-0.00} \big|_{\omega_M}
     {}\pm 0.02 \big|_{\omega_0}
     {}\pm 0.03 \big|_{f_B}
     {} \pm (?) \big|_{\rm syst.} \,.
\cr \nonumber
\eeq

\begin{figure}[t!ph]
\begin{center}
(a)
\psfrag{x}{\hspace{-1.7em}\footnotesize \parbox{6.5em}{\vspace{0.5em}$(n_+p')$ (GeV)} }
\psfrag{y}{\hspace{-3.0em}
 \parbox{2em}{$\frac{\xi_\pi(n_+p')}{\xi_\pi(m_B)}$} }
\hspace{-2em}
\includegraphics[height=0.27\textwidth]{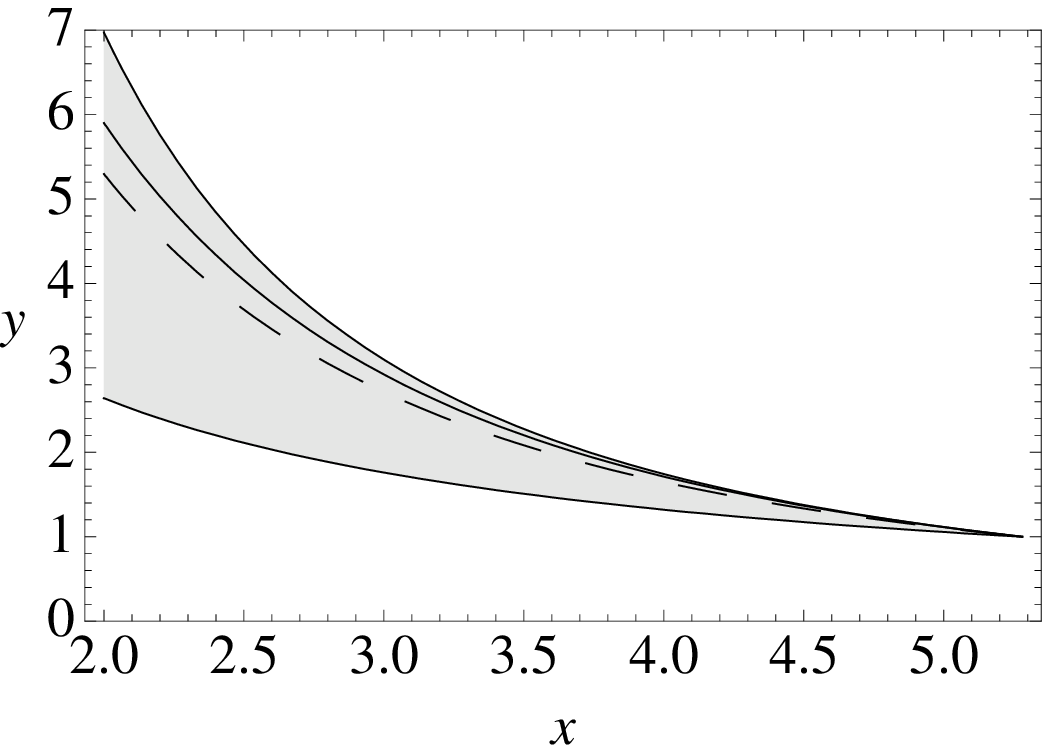}
\hspace{4em}
(b)
\hspace{-2em}
\psfrag{x}{\hspace{-2.2em}\footnotesize \parbox{7.5em}{\vspace{0.5em}$1/m_Q$ (GeV$^{-1}$)} }
\psfrag{y}{\hspace{-3.0em} \footnotesize \parbox{2em}{$R_\pi(m_Q)$} }
\includegraphics[height=0.27\textwidth]{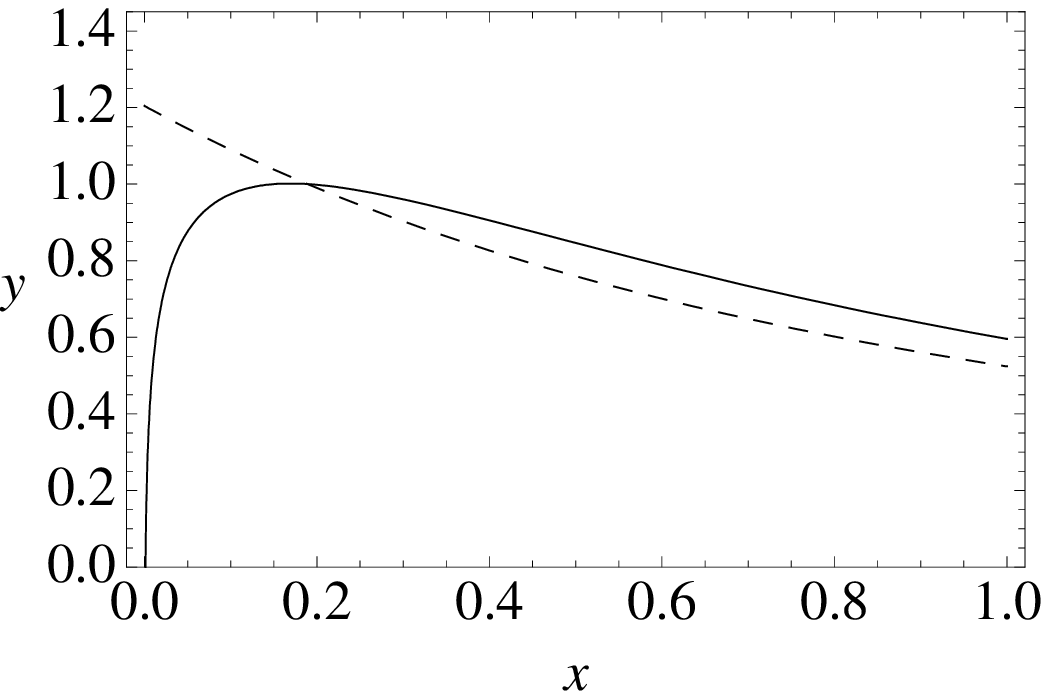}
\end{center}
\centerline{\parbox{0.95\textwidth}{\caption{\small (a) Energy dependence of the soft form factor
$\xi_\pi(n_+p')/\xi_\pi(m_B)$.
The grey band illustrates the range between a pure $1/(n_+p')^2$
and a pure $1/(n_+p')$ behaviour.
(b)
The $m_Q$ dependence of the soft form factor $R_\pi(m_Q)$,
(defined analogously as (\ref{eq:Rpar})),
as a function of $1/m_Q$. The solid line denotes
the NLO result, the dashed line the LO result.}
\label{fig:endeppi}}}
\end{figure}

In Fig.~\ref{fig:endeppi} we
show the energy and heavy-quark mass dependence of $\xi_\pi$
which is similar to that of $\xi_\perp$.

%%%%%%%%%%%%%%%%%%%%%%%%%%%%%%%%%%%%%%%%%%%%%%%%%%%%%%%

\subsection{Dependence on the shape of $\phi_B^-(\omega)$}

\label{modelphi}

\begin{figure}[t!ph]
\begin{center}
\psfrag{x}{\hspace{-1.7em}\footnotesize \parbox{6.5em}{\vspace{0.5em}$\omega$ (GeV)} }
\psfrag{y}{\hspace{-1.7em}
 \parbox{2em}{$ \phi_B^-$}}
(a)
\includegraphics[height=0.27\textwidth]{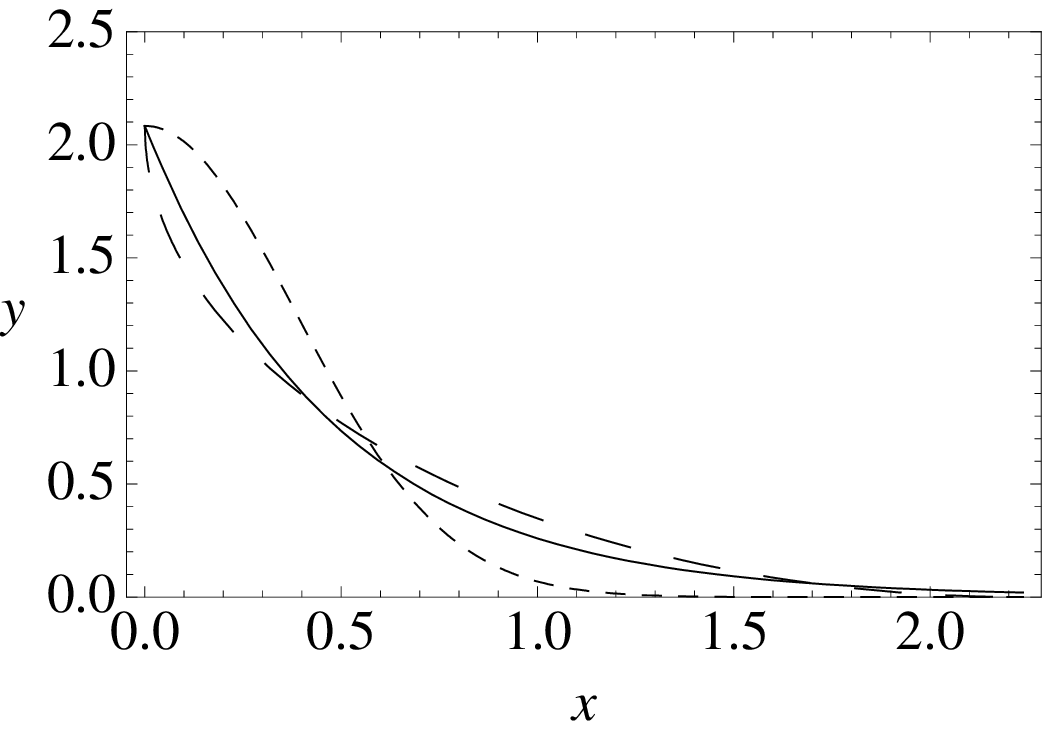}
\hspace{4em}
(b)
\psfrag{x}{\hspace{-2.2em}\footnotesize \parbox{7.5em}{\vspace{0.5em}$\omega_M$ (GeV)} }
\psfrag{y}{\hspace{-0.9em} \footnotesize \parbox{2em}{$\hat\xi_\parallel$} }
\hspace{-2em}
\includegraphics[height=0.27\textwidth]{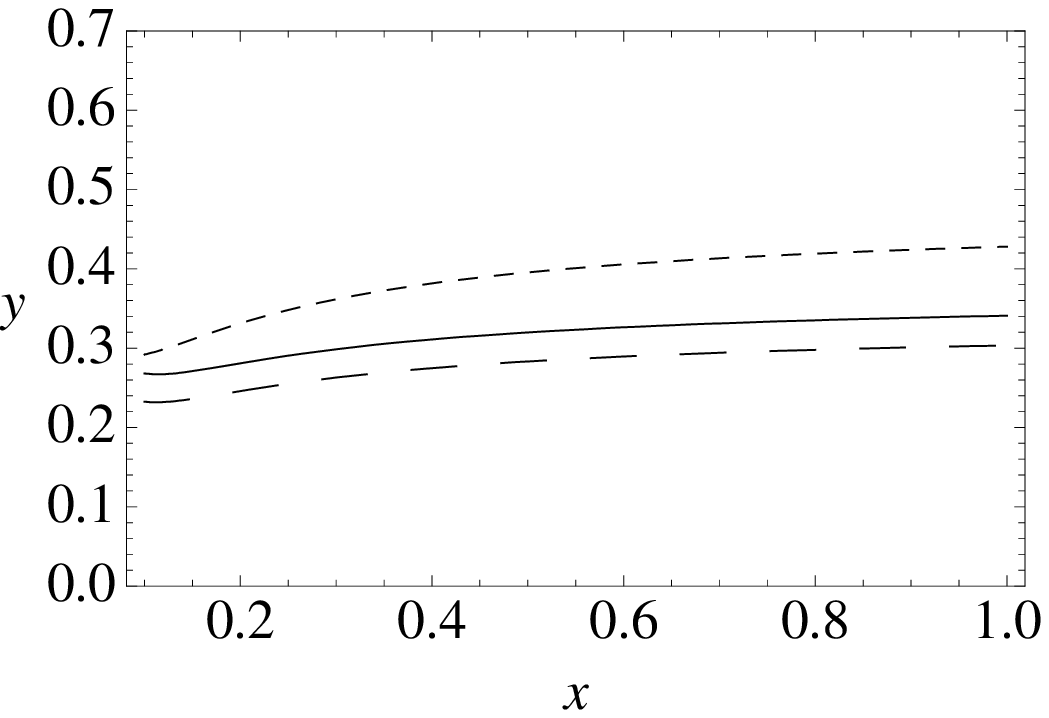}
\end{center}
\centerline{\parbox{0.95\textwidth}{\caption{\small (a) Three models for $\phi_B^-(\omega)$. Solid line: default model
(\ref{phidef}), short-dashed line (\ref{phi1}), long-dashed line (\ref{phi2}).
(b)
The resulting prediction for $\hat\xi_\parallel(m_B)$ as a function
of the Borel parameter ($\omega_s=0.4~$GeV, $\mu=1$~GeV).}
\label{fig:phimodel}}} 
\end{figure}

So far, in our error treatment, we have varied the
parameter $\omega_0 = 1/\phi_B^-(0)$ in a conservative range
in order to estimate the uncertainties related to the $B$\/-meson
LCDA. It should be stressed, however, that our result also depends
on the \emph{shape} of $\phi_B^-(\omega)$ in an essential way.
To illustrate this effect, we consider two alternative
parametrizations for $\phi_B^-(\omega)$
\beq
  \mbox{1:}
\qquad \phi_B^-(\omega) &=&
    \frac{1}{\omega_0} \, \exp\left[-\left(\frac{\omega}{\omega_1}\right)^2\right]
 \,, \qquad \omega_1 = \frac{2 \, \omega_0}{\sqrt \pi} \,;
\label{phi1}
\\[0.2em]
  \mbox{2:}\qquad \phi_B^-(\omega) &=&
    \frac{1}{\omega_0}
  \left( 1-\sqrt{\bigg(2-\frac{\omega}{\omega_2}\bigg) \frac{\omega}{\omega_2}}
   \, \right) \theta(\omega_2-\omega)
 \,, \qquad \omega_2 = \frac{4\, \omega_0}{4-\pi} \,.
\label{phi2}
\eeq
Both models have the same value at $\omega=0$ as our default model
(\ref{phidef}) and are normalized to unity (the radiative tail
\cite{Lee:2005gz}
of $\phi_B^-(\omega)$ is unimportant for our considerations,
because the sum rule focuses on small values of $\omega$). But the derivative
at $\omega=0$ takes extreme (but physically not excluded)
values $0$ and $\infty$, respectively.
In Fig.~\ref{fig:phimodel} we compare the three models for $\phi_B^-$
as well as the resulting soft form factor $\hat\xi_\parallel$ at
maximal recoil.

We observe that the effects on the soft form
factor are quite sizeable, ranging from about -10\% to +25\%
variations. In order to reduce the associated uncertainty,
one clearly needs additional qualitative and quantitative
information on the LCDA $\phi_B^-(\omega,\mu)$. In any case,
we have to conclude that the sum rule for the soft $B \to \pi$
form factor alone cannot be used to put stringent constraints
on $\phi_B^-(0)$.\footnote{In \cite{Khodjamirian:2005ea} 
independent information on
the soft $B \to \pi$ form factor was used to
determine the $1/\omega$ moment $\lambda_B$ of the LCDA $\phi_B^+(\omega)$,
by means of the approximate WW relation between $\phi_B^-(0)$ 
and $\lambda_B$ (see appendix~\ref{app:phiB}). 
In view of the large 
dependence on the exact shape of $\phi_B^-(\omega,\mu)$, such a
procedure appears questionable to us, at least, the
related uncertainty seems to
be significantly underestimated in \cite{Khodjamirian:2005ea}.
}

%%%%%%%%%%%%%%%%%%%%%%%%%%%%%%%%%%%%%%%%%%%%%%%%%%%%%%%

\subsection{Predictions for form factor ratios}

%%%%%%%%%%%%%%%%%%%%%%%%%%%%%%%%%%%%%%%%%%%%%%%%%%%%%%%%%

As discussed already in our first paper \cite{DeFazio:2005dx},
the numerical predictions for individual form factors are quite
sensitive to the hadronic input parameters related to the $B$\/-meson
distribution amplitude and the Borel and threshold parameters.
Certainly, part of these uncertainties should cancel in ratios
of form factors. In the following, we discuss two kind of ratios:
First, we consider ratios of soft form factors for $B \to \pi(K)$
and $B \to \rho(K^*)$ transitions.
Second, we consider the corrections to the symmetry relations
in the large-energy limit \cite{Charles:1998dr}, making use of
the sum-rule prediction for the factorizable corrections.

\subsubsection{Ratios of soft form factors}

\label{sec:ffratio}

\begin{table}[t!bp]
\centerline{\parbox{0.95\textwidth}{\caption{\label{tab:decrel} \small Ratios of normalized soft form factors
for original and modified sum rules (see text). The errors are obtained
by varying the sum-rule parameters (independently for every meson)
and $\omega_0$, added in quadrature.
For the original sum rule the errors are dominated by
the sum rule parameters. For the modified sum rule the errors
for the form factor ratios are dominated by the systematic error
related to the assumed correlation between the sum rules for the
soft form factor and for the decay constants, respectively. \label{tab:ratios}}}}
\begin{center}
\begin{tabular}{c || c | c }
\hline\hline
ratio: & $(\xi_\pi/f_\pi): (\hat \xi_\rho^\parallel/f_\rho^\parallel)$
       & $(\xi_\rho^\perp/f_\rho^\perp):(\hat \xi_\rho^\parallel/f_\rho^\parallel)$\\
\hline
&& \\[-0.8em]
original sum rule &  $1.18 {}^{+0.37}_{-0.32}$ &  $1.02 {}^{+0.28}_{-0.21}
$ \\[0.4em]
modified sum rule &  $1.05 {}^{+0.06}_{-0.04} {} \pm (?)_{\rm syst.}$ &  $0.87{}^{+0.06}_{-0.12}{} \pm (?)_{\rm syst.}$
\\[0.2em]
\hline\hline
\end{tabular}
\end{center}
\end{table}

It is convenient to normalize the soft form factors
to the corresponding decay constant and consider the
ratios
$$
   (\xi_\pi/f_\pi): (\hat \xi_\rho^\parallel/f_\rho^\parallel)
\qquad \mbox{and} \qquad
    (\xi_\rho^\perp/f_\rho^\perp):(\hat \xi_\rho^\parallel/f_\rho^\parallel) \,.
$$
In Table~\ref{tab:ratios} we present the numerical values obtained
from our predictions for the individual form factors at
$\mu=1.0$~GeV, obtained in the previous sections. The quoted
uncertainties are obtained by varying independently the sum rule
parameters for the two considered form factors, together with the
$\omega_0$ dependence (the dependence on $f_B$ drops
out). Using the original sum rules (\ref{sumpartree},\ref{sumperptree})
with (\ref{eq:phieff}),
the so-obtained uncertainty is dominated by the sum-rule
parameters and amounts to about 30\%.

Using, on the other hand, the modified sum rules
(\ref{eq:phieff_neu},\ref{eq:phieffp_neu},\ref{eq:phieffpi_neu})
(i.e.\ assuming a naive correlation between sum rule parameters
for soft form factors and decay constants), one gets very stable
result close to one, i.e.\  to the simple approximate relations
\beq \xi_P/f_P \simeq \hat \xi_V^\parallel/f_V^\parallel \simeq
\xi_V^\perp/f_V^\perp \label{eq:simple} \eeq which hold in this
case, with a parametric uncertainty of about 10-15\%. As already
said, this error does not include a systematic uncertainty which
cannot be estimated in a model-independent way.

\paragraph{$SU(3)$ flavour symmetry corrections:}

The approximate relations (\ref{eq:simple}) can directly be generalized to $B \to K$ and
$B \to K^*$ form factors, yielding an estimate for the $SU(3)$ ratios
\beq
&&  \xi_{K}/\xi_\pi  \simeq  f_K/f_\pi \qquad \mbox{and} \qquad
  \xi_{K^*}^{\parallel,\perp}/\xi_{\rho}^{\parallel,\perp}  \simeq
      f_{K^*}^{\parallel,\perp}/f_\rho^{\parallel,\perp} \,.
\eeq
Notice that the latter ratio is of particular importance for the
extraction of the CKM parameter $|V_{ts}/V_{td}|$ from the ratio of
the $B \to K^*\gamma$ and $B \to \rho \gamma$ branching fractions
\cite{Bosch:2001gv,Ali:2001ez,Beneke:2004dp,Ball:2006fw}.
The central values for the $SU(3)$ ratios,
$$
  \xi_{K^*}^\perp/ \xi_{\rho}^{\perp} \approx 1.1-1.2
$$
etc., are thus in line with the naive expectations.
However, it is to be stressed that the
theoretical uncertainty, to be assigned to the form-factor ratios,
depends very strongly on the assumptions about the sum-rule parameters.
Of course, we expect a {\it certain}\/ correlation between the values
for $s_0$ and $\omega_M$ in different sum rules. The way how to quantify
these correlations (and thus decreasing the uncertainties)
appears to us more controversial.

%%%%%%%%%%%%%%%%%%%%%%%%%%%%%%%%%%%%%%%%%%%%%%%%%%%%%%%%%%%%%%%%%%%%%%%%%

\subsubsection{Corrections to symmetry relations}

%%%%%%%%%%%%%%%%%%%%%%%%%%%%%%%%%%%%%%%%%%%%%%%%%%%%%%%%%%%%%%%%%%%%%%%%%

\begin{figure}
\begin{center}
\psfrag{x}{\hspace{-1.7em}\footnotesize \parbox{6.5em}{\vspace{0.5em}$\omega_M$ (GeV)} }
\psfrag{y}{\hspace{-2.9em}
 \parbox{2em}{$\frac{\Delta F_\parallel}{{\rm QCDF}}$} }
(a)
\includegraphics[height=0.27\textwidth]{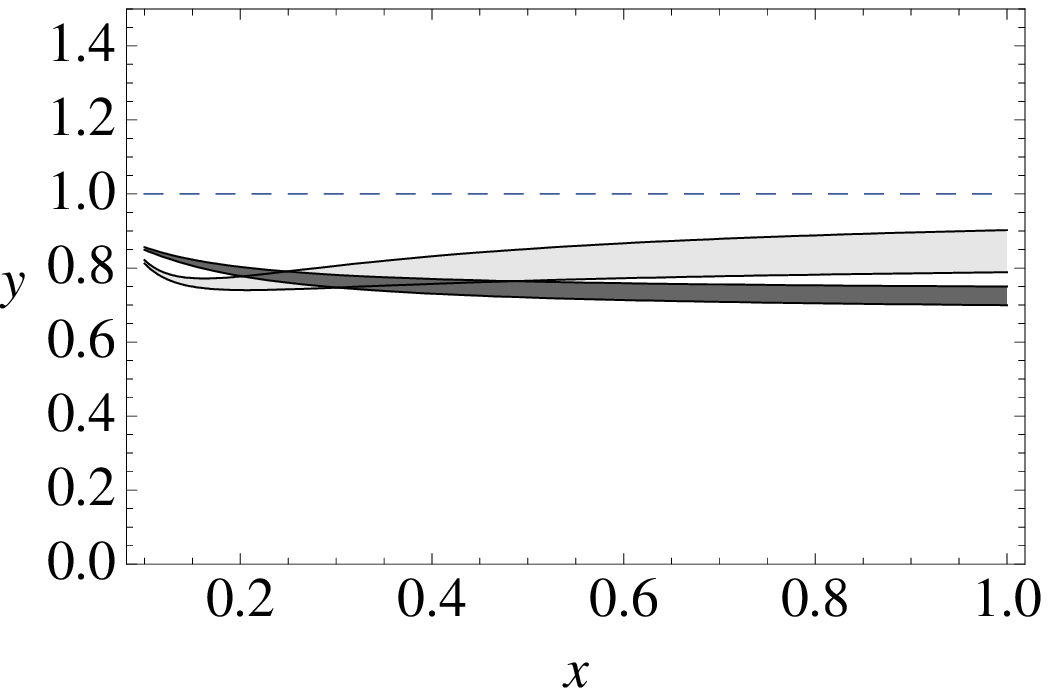}
\hspace{4em}
\psfrag{y}{\hspace{-2.9em}
 \parbox{2em}{$\frac{\Delta F_\perp}{{\rm QCDF}}$} }
(b)
\hspace{-2em}
\includegraphics[height=0.27\textwidth]{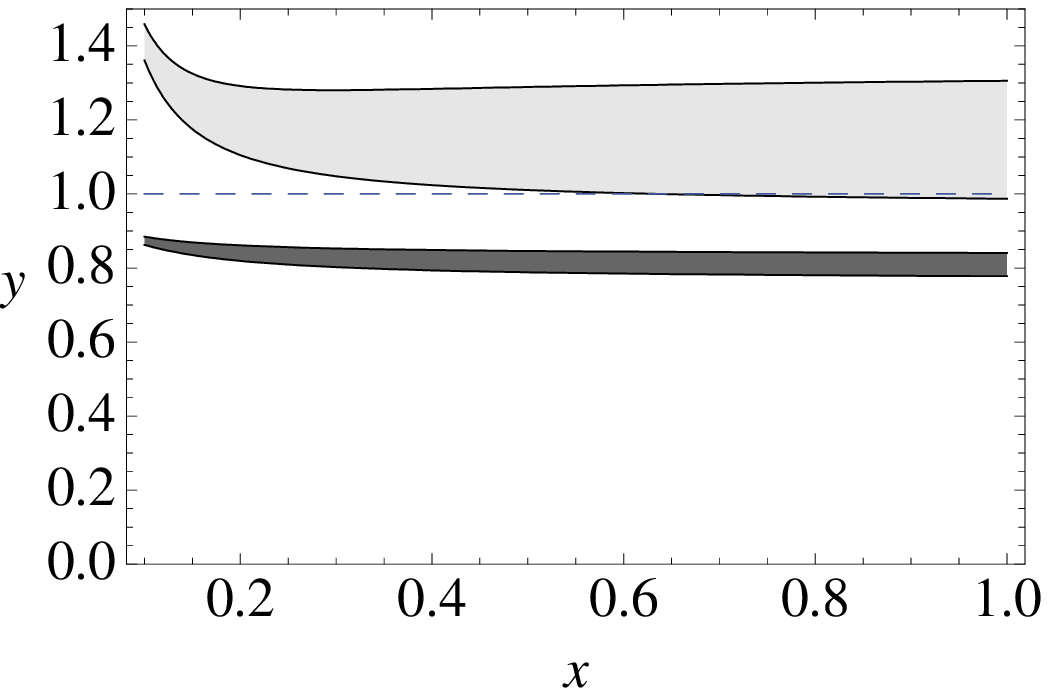}
\end{center}
\centerline{\parbox{0.95\textwidth}{\caption{\small (a) $\Delta F_\parallel$ from (\ref{eq:DeltaF1} -- light
gray band) or (\ref{eq:DeltaF2} -- dark gray band) normalized to the QCDF
result (\ref{eq:DeltaF3}) as a function of $\omega_M$ for
$\omega_s \in [0.35,0.45]$~GeV and $\mu=1.5$~GeV.
(b) The same for $\Delta F_\perp$ with $\omega_s \in [0.2,0.3]$~GeV.}
\label{fig:DeltaFparperp}}}
\end{figure}

\begin{figure}
\begin{center}
\psfrag{x}{\hspace{-1.7em}\footnotesize \parbox{6.5em}{\vspace{0.5em}$\omega_M$ (GeV)} }
\psfrag{y}{\hspace{-2.9em}
 \parbox{2em}{$\frac{\Delta F_\pi}{{\rm QCDF}}$} }
\includegraphics[height=0.27\textwidth]{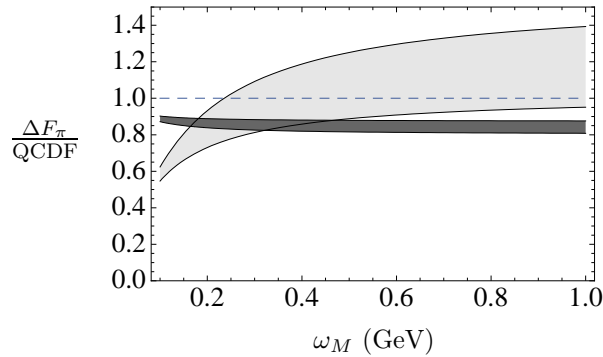}
\end{center}
\centerline{\parbox{0.95\textwidth}{\caption{\small $\Delta F_\pi$ from (\ref{eq:DeltaF1} -- light
gray band) or (\ref{eq:DeltaF2} -- dark grey band) normalized to the QCDF
result (\ref{eq:DeltaF3}) as a function of $\omega_M$ for
$\omega_s \in [0.15,0.25]$~GeV and $\mu=1.5$~GeV.} \label{fig:DeltaFpi}}}
\end{figure}

In Figs.~\ref{fig:DeltaFparperp} and \ref{fig:DeltaFpi}
we present a quantitative study of the factorizable form-factor
contributions, parametrized by the quantities
$$
    \Delta F_\parallel \,, \qquad
  \Delta F_\perp \,,\qquad\Delta F_\pi \,.
$$
We find it convenient to normalize to the QCD-factorization result
(\ref{eq:DeltaF3}) and compare the original sum rule (\ref{eq:DeltaF1})
and the modified one (\ref{eq:DeltaF2}). The threshold parameters
$\omega_s$ for the three cases are varied within the same ranges
we used for the soft form factor analysis, the factorization scale
is chosen as $\mu=1.5$~GeV, and the variation with the Borel
parameter $\omega_M$ is plotted.
The following observations can be made:
\begin{itemize}
  \item The modified sum rule shows a rather mild $\omega_M$
        dependence, and is typically 15-25\% smaller than the
        QCD factorization result (with asymptotic light-meson DA).
 
 \item The original sum rule shows a larger $\omega_M$
        dependence, and typically lies closer to or even above the
        QCD factorization result.

  \item In the case of transverse vector mesons, we observe a
        rather large deviation between the original sum rule
        (\ref{eq:DeltaF1}) and the modified one (\ref{eq:DeltaF2}).
        This discrepancy has also been seen in the soft form factor case.

\end{itemize}
In any case, the sum rule and the QCDF result are in qualitative
agreement. In particular, in the sum-rule approach we find no
potential source of anomalous enhancement of the $\Delta F_i$ as
it is sometimes required in phenomenological fits (see, for
instance, \cite{Bauer:2006iv} and references therein). For a more
precise quantitative statement, we would also have to consider
$\alpha_s^2$ contributions to the correlation function $\Pi_1$, as
well as include the $\alpha_s$ and $1/M^2$ corrections in the sum
rules for the decay constants, which is beyond the scope
of this work.

\subsection{Comparison with other work}

A study of $B \to \pi(K)$ and
$B \to \rho(K^*)$ form factors in the framework of light-cone sum rules
involving the $B$\/-meson distribution amplitudes also appeared in
\cite{Khodjamirian:2006st}. The authors use the same set-up with
the $B$\/-meson treated as an on-shell state in the heavy-mass limit of
QCD (which at tree-level coincides with HQET), and the
light mesons interpolated by appropriate currents. The correlation
functions are calculated only at tree level, but
contributions from 3-particle Fock states in the $B$\/-meson
(which have been neglected in our analysis) are taken into account.
The tree-level result for 2-particle contributions
to the sum rule coincides with ours for decays into light pseudoscalars
and longitudinally polarized vector mesons. In case of transversely
polarized vector mesons we used an interpolating tensor current, while
the authors of \cite{Khodjamirian:2006st} used a vector current, and
naturally the results for the corresponding sum rules look different. As
discussed above, a disadvantage of the tensor current is the ``pollution''
by the contribution of a near-lying axial-vector resonance, which is
avoided by taking the vector current in \cite{Khodjamirian:2006st}.
On the other hand, the correlation function with the tensor current is
more ``natural'' from the SCET point of view, since the vector-current
correlator turns out to be $\Lambda/m_b$\/-suppressed in the transverse case.
Therefore -- in contrast to \cite{Khodjamirian:2006st} --
our sum rule for transversely polarized vector mesons
does not receive leading contributions from the 3-particle Fock states at tree-level
and takes the same form as in the longitudinal or pseudoscalar case.
As a consequence, the leading dependence on the $B$\/-meson distribution
amplitude drops out in the form-factor ratios considered in Table~\ref{tab:decrel}.

Certainly, it will be desirable to combine the radiative corrections
with the 3-particle contributions to get a more reliable error estimate
for the physical $B \to M$ form factors.
This includes both, the elimination of the scale ambiguity
from $\phi_B^-(\omega,\mu)$ in the tree-level result through
$\alpha_s$ corrections to SCET correlation functions, and the effect
of $\Lambda/m_b$ corrections from sub-leading currents
and interactions in the SCET Lagrangian.
These issues will be left for future work.

\section{Summary}\label{concl}

In this paper we have studied light-cone sum rules
for heavy-to-light form factors
within the framework of soft-collinear
effective theory (SCET), generalizing our results for
the $B \to \pi$ form factors in \cite{DeFazio:2005dx}
to the cases of decays into light vector mesons.

The conceptual advantage of our formalism -- compared
to the more traditional light-cone sum rules in QCD
(see e.g.\ \cite{Ball:1998kk,Khodjamirian,Ball:1998tj,Colangelo:2000dp})
-- is a clear separation of factorizable ("soft")
and non-factorizable ("hard") contributions, on the basis
of a systematic expansion in inverse powers of the $b$\/-quark
mass, already on the level of the correlation functions.
This is particularly
useful to make contact to the QCD factorization approach, both
on the qualitative and on the quantitative level.
In the SCET correlation functions, involving
interpolating currents for the light mesons under consideration,
the hard-collinear dynamics is factorized from the soft dynamics.
The latter is described by non-perturbative light-cone wave functions for
the $B$ meson. This factorization has been explicitly
verified to order $\alpha_s$ accuracy (neglecting 3-particle
LCDAs), making use of a new result for the anomalous dimension
kernel $\gamma_-(\omega,\omega')$ obtained in \cite{Bell:2007}.
The hard dynamics from scales of order $m_b$ is already integrated
out, and appears in matching coefficients of the decay currents
in SCET.

Sum rules for the heavy-to-light form factors are, as usual,
obtained from a dispersive analysis of the correlation function,
introducing non-perturbative parameters associated to
the continuum threshold and the Borel transformation.
Besides the light-cone distribution amplitude of the $B$\/-meson
itself, these parameters represent the main source of theoretical
(systematic) uncertainties in our numerical predictions for the
form factors.
In this context, it is important to realize the different
dynamical nature of factorizable and
non-factorizable form factor contributions, which also reflects
itself in the respective sum rules. Considering the sum rule for the
factorizable form factor (\ref{eq:DeltaF1}) in the heavy-quark mass limit,
the dependence on the sum-rule parameters factorizes from
the soft convolution integral involving the $B$\/-meson
distribution amplitude. In this case, the sum-rule parameters
can be viewed as universal properties of the spectrum of the
interpolating current, and by comparing with the sum rule for
the meson decay constants, one formally recovers the QCD factorization
result.

In contrast, the analogous limit for
the non-factorizable form factor contributions (\ref{eq:phieff}) does
not exist beyond tree level, and the responsible non-factorizable
logarithmic dependence on the threshold parameter has the same
physical origin as the endpoint divergences observed in the
standard QCD factorization approach.
As a consequence, the sum rule parameters in (\ref{eq:phieff}) are to
be considered as independent non-perturbative input, specific
to the soft (i.e.\ endpoint-dominated) dynamics. (In terms of
the interpolating currents in SCET, see (\ref{eq:Jpar}) and (\ref{eq:Jperp}),
the factorizable terms probe the first terms with two hard-collinear
fields, whereas the non-factorizable terms probe the terms with
one hard-collinear and one soft field.)
Following a conservative treatment of theoretical systematics,
our quantitative estimates for the soft form factor contributions
thus leads to rather large uncertainties, where the accepted
ranges for the sum rule parameters are constrained by requiring
sufficient stability under variations of the Borel parameter,
and a sufficiently small continuum pollution. The resulting
predictions for the soft form factors are in general agreement
with the expectations from QCD light-cone sum rules and
QCD factorization. The NLO corrections, calculated in this paper,
are numerically significant
and can reduce the tree-level estimates by up to 30\%.
Concerning the energy-dependence of the
soft form factors, we find a significant deviation of the
simple $1/E^2$ behaviour, in particular for longitudinally polarized
vector mesons.

The uncertainties related to the sum-rule parameters and
the $B$\/-meson distribution amplitudes can be somewhat
reduced by considering form factor \emph{ratios}. Still,
a proper error estimate requires to make certain assumptions
about correlations between the threshold and Borel parameters
in different sum rules. In the simplest case, we find that to first
approximation the soft form factors scale with the respective
light meson's decay constant.

\section*{Acknowledgements}

We thank Pietro Colangelo, Alex Khodjamirian, Thomas Mannel and Nils Offen
for helpful discussions, and Guido Bell
for valuable comments on the manuscript. 
F.DF.\ and T.F.\ are grateful to Andrzej Buras and his group
at the Physics Department of the Technical University in 
Munich for the warm hospitality, and the Cluster of Excellence
``Origin and Structure of the Universe'' for financial support.
This work was
supported in part by the EU contract No.~MRTN-CT-2006-035482,
"FLAVIAnet". T.F.\ is supported by the German Ministry of Research
(BMBF, contract No.~05HT6PSA).

%\clearpage

%\clearpage

\begin{appendix}

\section{Calculation of the imaginary part at one-loop}

\label{app:impart}

The imaginary part from diagrams (a1-a4),
after $\overline{\rm MS}$ renormalization, determines
the one-loop contribution to the 
jet function in inclusive $b\to u$ decays
\cite{Bauer:2003pi,Bosch:2004th},
\beq
  \frac{1}{\pi} \, {\rm Im}[\Pi_{\parallel,\perp}^a] &=&
  \frac{\alpha_s C_F}{4\pi} \, f_B m_B \, \int\limits_0^\infty d\omega \,
  \phi_B^-(\omega)
\cr && {} \times
  \left\{
 (7-\pi^2) \, \delta(u) -3 \left(\frac{1}{u}\right)_*^{[\mu^2/n_+p']}
  +4 \left(\frac{\ln[u (n_+p')/\mu^2]}{u}\right)_*^{[\mu^2/n_+p']}
  \right\} \,,
\label{jet}
\eeq
where $u=(n_-p - \omega + i\eta)$, 
and the modified plus-distributions are defined through
(see e.g.~\cite{DeFazio:1999sv}),
\beq
\int\limits_{\leq 0}^M du \, F(u)\left(\frac{1}{u}\right)_*^{[m]}&=&
\int\limits_0^M du\frac{F(u)-F(0)}{u}+F(0)\ln\biggl(\frac{M}{m}\biggr),
\\
\int\limits_{\leq 0}^M du \, F(u)
\left(\frac{\ln(u/m)}{u}\right)_*^{[m]}&=&
\int\limits_0^M du\frac{F(u)-F(0)}{u}\ln\frac{u}{m}+\frac{F(0)}{2}
\ln^2\biggl(\frac{M}{m}\biggr)\,.
\eeq
Using (\ref{eq:Borel}),
taking into account the continuum subtraction
$\omega' < \omega_s$, and performing the $\omega$ integration
in (\ref{jet}),
this contributes to the sum rule as
\beq
  \hat B [\Pi_{\parallel,\perp}^a](\omega_M)- \mbox{cont.}
&=&
 \frac{ f_B m_B }{\omega_M}
 \, \frac{\alpha_s C_F}{4\pi}\, \int\limits_0^{\omega_s} d\omega'
\, e^{-\omega'/\omega_M} \,
\cr &&
{} \times
  \left\{
 \left(
 7-\pi^2
 + 3 \ln \left[\frac{\mu^2}{\omega'(n_+p')}\right]
 + 2 \ln^2 \left[\frac{\mu^2}{\omega'(n_+p')}\right]
\right) \phi_B^-(\omega')
\right.
\cr &&
\left. \quad
{} + \int\limits_0^{\omega'} d\omega
 \left( 4 \ln \left[\frac{\mu^2}{(\omega'-\omega) (n_+p')}\right] + 3 \right)
 \, \frac{\phi_B^-(\omega')-\phi_B^-(\omega)}{\omega'-\omega}
  \right\}
\cr &&
\eeq

For convenience, we summarize the results for the imaginary part
from the basic structures appearing in the one-loop correlation function:
\beq
\frac{1}{\pi} {\rm Im} \, \int\limits_0^\infty
  \frac{d\omega}{\omega-\omega'} \, L(\mu) \, \phi_B^-(\omega)
&=&
\int\limits_0^{\omega'} \, d\omega \,
 \frac{\phi_B^-(\omega)-\phi_B^-(\omega')}{\omega-\omega'}
+ L_0 \,  \phi_B^-(\omega')
\ ,
\\
\frac{1}{\pi} {\rm Im} \, \int\limits_0^\infty
  \frac{d\omega}{\omega-\omega'} \left[L(\mu) \right]^2 \phi_B^-(\omega)
&=&
\int\limits_0^{\omega'} \, d\omega \,
  \left(2L_0 - 2L_1\right)
  \frac{\phi_B^-(\omega)-\phi_B^-(\omega')}{\omega-\omega'}
+ \left(L_0^2-\frac{\pi^2}{3}\right) \phi_B^-(\omega')
\, ,
\cr &&
\\
 \frac{1}{\pi} {\rm Im} \, \int\limits_0^\infty
  \frac{d\omega}{\omega} \, L_1 \, \phi_B^-(\omega)
&=&
- \int\limits_{\omega'}^\infty \, d\omega \, \ln \frac{\omega}{\omega'}
 \left(\frac{d}{d\omega} \phi_B^-(\omega)\right)
\ ,
\\
\frac{1}{\pi} {\rm Im} \, \int\limits_0^\infty
  \frac{d\omega}{\omega-\omega'} \, L_1\, \phi_B^-(\omega)
&=&
- \int\limits_{\omega'}^\infty d\omega \,
  \ln \left[\frac{\omega}{\omega'}-1\right]
  \left(\frac{d}{d\omega} \phi_B^-(\omega)\right) \ ,
\\
 \frac{1}{\pi} {\rm Im} \, \int\limits_0^\infty
  \frac{d\omega}{\omega-\omega'}
\, \left[L_1\right]^2  \phi_B^-(\omega)
&=&
- \int\limits_{\omega'}^\infty d\omega \,
  \ln^2 \left[\frac{\omega}{\omega'}-1\right]
  \left(\frac{d}{d\omega} \phi_B^-(\omega)\right)
 - \frac{\pi^2}{3} \, \phi_B^-(\omega')
\ ,
\cr &&
\eeq
and
\beq
&&
 \frac{1}{\pi} {\rm Im} \, \int\limits_0^\infty
  \frac{d\omega}{\omega-\omega'} \, L_1 \, L(\mu)
\,  \phi_B^-(\omega)
=\int\limits_0^{\omega'} d\omega \, L_1
   \, \frac{\phi_B^-(\omega)-\phi_B^-(\omega')}{\omega-\omega'}
\cr && \qquad {}
- \int\limits_{\omega'}^\infty d\omega \,
  \left(L_0 \, \ln \left[\frac{\omega}{\omega'}-1\right] -\frac12 \,
      \ln^2 \left[\frac{\omega}{\omega'}-1\right] - \frac{\pi^2}{6} \right)
  \left(\frac{d}{d\omega} \phi_B^-(\omega)\right)
 \, .
\cr &&
\eeq
where
$$
  L(\mu)=\ln \left[-\frac{\mu^2}{(\omega'-\omega)\, n_+ p} \right] \ ,
\qquad
  L_1 =
        \ln \left[1 - \frac{\omega}{\omega'} \right] \ ,
\qquad
  L_0 = \ln \left[\frac{\mu^2}{\omega' \, n_+p} \right] \ ,
$$
and ${\rm Im}[w]<0$.
Notice that the imaginary part resulting from the diagrams (b1+b2)
also has support for $\omega > \omega'$.

\section{$B$\/-meson LCDAs in $D\neq 4$ dimensions}

\label{app:phiB}

We follow the derivation as given in the appendix
of \cite{Beneke:2000wa}.
Starting point is the general decomposition of the two-particle
$B$\/-meson matrix element \cite{Grozin:1996pq}
\begin{equation}
\langle 0 | \bar q_\beta (z) \, P(z,0) \, b_\alpha(0) |
  \bar B(p) \rangle = - \frac{i f_B M}{4}
  \left[ \frac{1+\slash v}{2} \,
         \left\{ 2 \tilde \phi_B^+(t,z^2) +
                 \frac{\tilde \phi_B^-(t,z^2)-\tilde \phi_B^+(t,z^2)}{t}
                 \, \slash z \right\} \, \gamma_5 \right]_{\alpha\beta}\,,
\label{ansatz1}
\end{equation}
where $v\cdot z=t$ and, for the moment, we allowed the separation between
the quark fields to be off the light-cone, $z^2 \neq 0$. $P(z,0)$ is the usual
path-ordered exponential which, however, will be neglected in the derivation
of the Wandzura-Wilczek (WW) relation.

The equation of motion for the light
spectator quark relates $\tilde \phi_B^-(l_+)$ to $\tilde\phi_B^+(l_+)$
in the approximation that the three-particle
amplitudes are set to zero (the effect of three-particle LCDAs
in $D\neq 4$ is discussed in \cite{Bell:2007}).
We require the right-hand side of  Eq.~(\ref{ansatz1}) to vanish after
application of $\left[\slash \partial_{z^2}\right]_{\beta\gamma}$
(which is true only if the three-particle Fock-state $b\bar q g$
is neglected), and $\tilde \phi^B_\pm(t,z^2)$
not to vanish as $z^2\to 0$. Writing
$$
 \partial_z^\mu = v^\mu \, \partial_t + 2 z^{\mu} \, \partial_{z^2} \,,
$$
 we obtain two constraints from the two
independent Dirac structures $\frac{1+\slash v}{2} \gamma_5$ and
$\frac{1+\slash v}{2} \gamma_5 \slash {z}$
\begin{eqnarray}
  \partial_t \tilde \phi_B^-
 + \frac{D-2}{2t} \, (\tilde\phi_B^- - \tilde\phi_B^+) \, \Big|_{z^2=0} &=&0
\label{ww1}
\\
 - \partial_t (\tilde\phi_B^- - \tilde\phi_B^+) + 4 t \, \partial_{z^2} \tilde\phi_B^+
 + \frac{1}{t} \, (\tilde\phi_B^- - \tilde\phi_B^+) \, \Big|_{z^2=0} &=&0 \,.
\label{wwaux}
\end{eqnarray}
The first relation modifies the WW relation as quoted in (110) of
\cite{Beneke:2000wa}.
Inserting the $t$\/-derivative of the first equation into the second
one, the latter can be simplified as
\begin{eqnarray}
  \frac{\partial \tilde \phi_B^+}{\partial z^2} +
    \frac{1}{2(D-2)} \, \frac{\partial^2 \tilde \phi_B^-}{\partial t^2}
    \,  \Big|_{z^2=0} &=& 0 \,,
\end{eqnarray}
which modifies equation~(111) in \cite{Beneke:2000wa}.
The momentum-space representation of (\ref{ww1}) now reads
\begin{equation}
\label{brel1}
\int_0^{l_+} d\eta \,\left(\phi_B^-(\eta)-
\phi_B^+(\eta)\right) = \frac{2}{D-2} \, l_+ \phi_B^-(l_+)
\qquad\mbox{or}\qquad
 \phi_B^+(l_+) = - \frac{2}{D-2} \, l_+  \, \phi^{\prime\,-}_B(l_+),
\end{equation}
which is solved by
\begin{equation}
\phi_B^-(l_+) = \frac{D-2}{2} \, \int\limits_0^1
                    \frac{d\eta}{\eta}\,\phi_B^+(l_+/\eta).
\label{ww2}
\end{equation}
The factor $2/(D-2)$ also appears in the momentum-space projector
for the $B$\/-meson wave function in the WW approximation
\begin{eqnarray}
\label{bproj}
M^B_{\beta\alpha} &=&
- \frac{i f_B M}{4}\Bigg[ \frac{1+\slash v}{2} \Bigg\{
\phi_B^+(\omega)\,\slash n_+ +  \phi_B^-(\omega)\,\slash n_-
- \frac{2}{D-2} \, \omega \phi_B^-(\omega) \,\gamma_\perp^\mu\frac{\partial}{\partial l_{\perp\mu}}
\Bigg\} \, \gamma_5 \Bigg]_{\alpha\beta}.
\end{eqnarray}
where the transverse Dirac matrices $\gamma_\perp$ obey $\gamma_\perp^\mu \gamma^\perp_\mu = D-2$.

\end{appendix}


\begin{thebibliography}{99}

%\cite{Beneke:2000wa}
\bibitem{Beneke:2000wa}
M.~Beneke and T.~Feldmann,
%``Symmetry-breaking corrections to heavy-to-light B meson form factors at
%large recoil,''
Nucl.\ Phys.\ B {\bf 592} (2001) 3
[hep-ph/0008255].
%%CITATION = HEP-PH 0008255;%%


%\cite{Beneke:2003pa}
\bibitem{Beneke:2003pa}
M.~Beneke and T.~Feldmann,
%``Factorization of heavy-to-light form factors in soft-collinear effective
%theory,''
Nucl.\ Phys.\ B {\bf 685} (2004) 249
[hep-ph/0311335].
%%CITATION = HEP-PH 0311335;%%



%\cite{Lange:2003pk}
\bibitem{Lange:2003pk}
%%CITATION = HEP-PH 0311345;%%
R.~J.~Hill and M.~Neubert,
%``Spectator interactions in soft-collinear effective theory. ((U)),''
Nucl.\ Phys.\ B {\bf 657} (2003) 229
[hep-ph/0211018];
%%CITATION = HEP-PH 0211018;%%
B.~O.~Lange and M.~Neubert,
 %``Factorization and the soft overlap contribution to heavy-to-light form
%factors,''
Nucl.\ Phys.\ B {\bf 690} (2004) 249
[hep-ph/0311345].



%\cite{Bauer:2002aj}
\bibitem{Bauer:2002aj}
C.~W.~Bauer, D.~Pirjol and I.~W.~Stewart,
%``Factorization and end-point singularities in heavy-to-light decays,''
Phys.\ Rev.\ D {\bf 67} (2003) 071502
[hep-ph/0211069].
%%CITATION = HEP-PH 0211069;%%


%\cite{DeFazio:2005dx}
\bibitem{DeFazio:2005dx}
  F.~De Fazio, T.~Feldmann and T.~Hurth,
  %``Light-cone sum rules in soft-collinear effective theory,''
  Nucl.\ Phys.\ B {\bf 733} (2006) 1
  [hep-ph/0504088].
  %%CITATION = HEP-PH 0504088;%%%%%\cite{Hurth:2005mm}

%\cite{Hurth:2005mm}
\bibitem{Hurth:2005mm}
F.~De Fazio, T.~Feldmann and T.~Hurth
%``The B --> pi form factor from light-cone sum rules in soft-collinear
%effective theory,''
PoS {\bf HEP2005} (2006) 215
[hep-ph/0509167].
%%CITATION = POSCI,HEP2005,215;%%


%\cite{Bauer:2000yr}
\bibitem{Bauer:2000yr}
C.~W.~Bauer, S.~Fleming, D.~Pirjol and I.~W.~Stewart,
%``An effective field theory for collinear and soft gluons: Heavy to light
%decays,''
Phys.\ Rev.\ D {\bf 63} (2001) 114020
[hep-ph/0011336].
%%CITATION = HEP-PH 0011336;%%
%\cite{Bauer:2001ct}
%\bibitem{Bauer:2001ct}
C.~W.~Bauer and I.~W.~Stewart,
%``Invariant operators in collinear effective theory,''
Phys.\ Lett.\ B {\bf 516} (2001) 134
[hep-ph/0107001].
%%CITATION = HEP-PH 0107001;%%


%\cite{Beneke:2002ph}
\bibitem{Beneke:2002ph}
M.~Beneke, A.~P.~Chapovsky, M.~Diehl and T.~Feldmann,
%``Soft-collinear effective theory and heavy-to-light currents beyond  leading
%power,''
Nucl.\ Phys.\ B {\bf 643} (2002) 431
[hep-ph/0206152];
%%CITATION = HEP-PH 0206152;%%
%\cite{Beneke:2002ni}
%\bibitem{Beneke:2002ni}
M.~Beneke and T.~Feldmann,
%``Multipole-expanded soft-collinear effective theory with non-abelian gauge
%symmetry,''
Phys.\ Lett.\ B {\bf 553} (2003) 267
[hep-ph/0211358].
%%CITATION = HEP-PH 0211358;%%


%\cite{Grozin:1996pq}
\bibitem{Grozin:1996pq}
A.~G.~Grozin and M.~Neubert,
%``Asymptotics of heavy-meson form factors,''
Phys.\ Rev.\ D {\bf 55} (1997) 272
[hep-ph/9607366].
%%CITATION = HEP-PH 9607366;%%


%\cite{Charles:1998dr}
\bibitem{Charles:1998dr}
J.~Charles, A.~Le Yaouanc, L.~Oliver, O.~Pene and J.~C.~Raynal,
%``Heavy-to-light form factors in the heavy mass to large energy limit of
%{QCD},''
Phys.\ Rev.\ D {\bf 60} (1999) 014001
[hep-ph/9812358].
%%CITATION = HEP-PH 9812358;%%


%\cite{Bauer:2003pi}
\bibitem{Bauer:2003pi}
C.~W.~Bauer and A.~V.~Manohar,
%``Shape function effects in B $\to$ X/s gamma and B $\to$ X/u l nu decays,''
Phys.\ Rev.\ D {\bf 70} (2004) 034024
[hep-ph/0312109].
%%CITATION = HEP-PH 0312109;%%

%\cite{Bosch:2004th}
\bibitem{Bosch:2004th}
S.~W.~Bosch, B.~O.~Lange, M.~Neubert and G.~Paz,
%``Factorization and shape-function effects in inclusive B-meson decays,''
Nucl.\ Phys.\ B {\bf 699} (2004) 335
[hep-ph/0402094].
%%CITATION = HEP-PH 0402094;%%


%\cite{Lange:2003ff}
\bibitem{Lange:2003ff}
B.~O.~Lange and M.~Neubert,
%``Renormalization-group evolution of the B-meson light-cone distribution
%amplitude,''
Phys.\ Rev.\ Lett.\  {\bf 91} (2003) 102001
[hep-ph/0303082].
%%CITATION = HEP-PH 0303082;%


%\cite{Bell:2007}
\bibitem{Bell:2007}
G.~Bell and T.~Feldmann,
%``Light-cone distribution amplitudes for non-relativistic bound states,''
arXiv:0711.4014 [hep-ph];
%%CITATION = ARXIV:0711.4014;%%
%
%``Modelling light-cone distribution amplitudes from non-relativistic bound states,''
arXiv:0802.2221 [hep-ph].
  %%CITATION = ARXIV:0802.2221;%%
%Siegen preprint SI-HEP-2007-20 (to be published).
%


%\cite{Bell:2005gw}
\bibitem{Bell:2005gw}
  G.~Bell and T.~Feldmann,
  %``Heavy-to-light form factors for non-relativistic bound states,''
  Nucl.\ Phys.\ Proc.\ Suppl.\  {\bf 164} (2007) 189
  [hep-ph/0509347].
  %%CITATION = NUPHZ,164,189;%%


\bibitem{Beneke:2005gs}
  M.~Beneke and D.~Yang,
  %``Heavy-to-light B meson form factors at large recoil energy: Spectator
  %scattering corrections,''
  Nucl.\ Phys.\  B {\bf 736} (2006) 34
  [hep-ph/0508250].
  %%CITATION = NUPHA,B736,34;%%


%\cite{Kirilin:2005xz}
\bibitem{Kirilin:2005xz}
  G.~G.~Kirilin,
  %``Loop corrections to the form factors in B --> pi l nu decay,''
  hep-ph/0508235.
  %%CITATION = HEP-PH/0508235;%%

%\cite{Becher:2004kk}
\bibitem{Becher:2004kk}
  T.~Becher and R.~J.~Hill,
  %``Loop corrections to heavy-to-light form factors and evanescent operators
  %in SCET,''
  JHEP {\bf 0410} (2004) 055
  [hep-ph/0408344].
  %%CITATION = JHEPA,0410,055;%%

%\cite{Lee:2005gz}
\bibitem{Lee:2005gz}
  S.~J.~Lee and M.~Neubert,
  %``Model-independent properties of the B-meson distribution amplitude,''
  Phys.\ Rev.\ D {\bf 72} (2005) 094028
  [hep-ph/0509350].
  %%CITATION = HEP-PH 0509350;%%

%\cite{Braun:2003wx}
\bibitem{Braun:2003wx}
V.~M.~Braun, D.~Y.~Ivanov and G.~P.~Korchemsky,
%``The B-meson distribution amplitude in QCD,''
Phys.\ Rev.\ D {\bf 69} (2004) 034014
[hep-ph/0309330].
%%CITATION = HEP-PH 0309330;%%


%\cite{Ball:1996tb}
\bibitem{Ball:1996tb}
P.~Ball and V.~M.~Braun,
%``The $\rho$ Meson Light-Cone Distribution Amplitudes of Leading Twist
%Revisited,''
Phys.\ Rev.\ D {\bf 54} (1996) 2182
[hep-ph/9602323].
%%CITATION = HEP-PH 9602323;%%


%\cite{Ball:1998kk}
\bibitem{Ball:1998kk}
  P.~Ball and V.~M.~Braun,
  %``Exclusive semileptonic and rare B meson decays in {QCD},''
  Phys.\ Rev.\  D {\bf 58} (1998) 094016
  [hep-ph/9805422]

%\cite{Ball:2004rg}
\bibitem{Ball:2004rg}
  P.~Ball and R.~Zwicky,
  %``B/(d,s) --> rho, omega, K*, Phi decay form factors from light-cone sum
  %rules revisited,''
  Phys.\ Rev.\  D {\bf 71} (2005) 014029
  [hep-ph/0412079].
  %%CITATION = PHRVA,D71,014029;%%

%\cite{Beneke:2004dp}
\bibitem{Beneke:2004dp}
  M.~Beneke, T.~Feldmann and D.~Seidel,
  %``Exclusive radiative and electroweak b --> d and b --> s penguin decays  at
  %NLO,''
  Eur.\ Phys.\ J.\ C {\bf 41} (2005) 173
  [hep-ph/0412400].
  %%CITATION = HEP-PH 0412400;%%


%\cite{Hill:2005ju}
\bibitem{Hill:2005ju}
  R.~J.~Hill,
  %``Heavy-to-light meson form factors at large recoil,''
  Phys.\ Rev.\  D {\bf 73} (2006) 014012
  [hep-ph/0505129].
  %%CITATION = PHRVA,D73,014012;%%



%\cite{Khodjamirian:2005ea}
\bibitem{Khodjamirian:2005ea}
  A.~Khodjamirian, T.~Mannel and N.~Offen,
  %``B-meson distribution amplitude from the B --> pi form factor,''
  Phys.\ Lett.\ B {\bf 620}, 52 (2005)
  [hep-ph/0504091].
%%CITATION = HEP-PH 0504091;%%



%\cite{Chay:2005gh}
\bibitem{Chay:2005gh}
  J.~Chay, C.~Kim and A.~K.~Leibovich,
  %``SCET sum rules for heavy-to-light form factors,''
  Phys.\ Lett.\  B {\bf 628} (2005) 57
  [hep-ph/0508157].
  %%CITATION = PHLTA,B628,57;%%

%\cite{Bosch:2001gv}
\bibitem{Bosch:2001gv}
  S.~W.~Bosch and G.~Buchalla,
  %``The radiative decays B --> V gamma at next-to-leading order in QCD,''
  Nucl.\ Phys.\ B {\bf 621} (2002) 459
  [hep-ph/0106081].
  %%CITATION = HEP-PH 0106081;%%

%\cite{Ali:2001ez}
\bibitem{Ali:2001ez}
  A.~Ali and A.~Y.~Parkhomenko,
  % ``Branching ratios for B --> rho gamma decays in next-to-leading order in
  %alpha(s) including hard spectator corrections,''
  Eur.\ Phys.\ J.\ C {\bf 23} (2002) 89
  [hep-ph/0105302].
  %%CITATION = HEP-PH 0105302;%%


%\cite{Ball:2006fw}
\bibitem{Ball:2006fw}
  P.~Ball and R.~Zwicky,
  %``B --> K* gamma vs B --> rho gamma and |V(td)/V(ts)|,''
  hep-ph/0608009.
  %%CITATION = HEP-PH 0608009;%%


\bibitem{Bauer:2006iv}
  C.~W.~Bauer,
  %``Hadronic B decays from SCET,''
  eConf {\bf C060409} (2006) 039
  [hep-ph/0606018].
  %%CITATION = HEP-PH 0606018;%%}


%\cite{Khodjamirian:2006st}
\bibitem{Khodjamirian:2006st}
  A.~Khodjamirian, T.~Mannel and N.~Offen,
  %``Form factors from light-cone sum rules with B-meson distribution
  %amplitudes,''
  Phys.\ Rev.\  D {\bf 75} (2007) 054013
  [hep-ph/0611193].
  %%CITATION = PHRVA,D75,054013;%%




%\cite{Khodjamirian}
\bibitem{Khodjamirian}
  A.~Khodjamirian and R.~R\"uckl,
  %``{QCD} sum rules for exclusive decays of heavy mesons,''
  Adv.\ Ser.\ Direct.\ High Energy Phys.\  {\bf 15} (1998) 345
  [hep-ph/9801443].
%%CITATION = HEP-PH 9801443;%%

%\cite{Ball:1998tj}
\bibitem{Ball:1998tj}
  %%CITATION = PHRVA,D58,094016;%%
P.~Ball,
%``B $\to$ pi and B $\to$ K transitions from {QCD} sum rules on the
%light-cone,''
JHEP {\bf 9809} (1998) 005
[hep-ph/9802394].
%%CITATION = HEP-PH 9802394;%%


%\cite{Colangelo:2000dp}
\bibitem{Colangelo:2000dp}
P.~Colangelo and A.~Khodjamirian,
%``QCD sum rules: A modern perspective,''
hep-ph/0010175.
%%CITATION = HEP-PH 0010175;%%


\bibitem{DeFazio:1999sv}
F.~De Fazio and M.~Neubert,
%``B $\to$ X/u l anti-nu/l decay distributions to order alpha(s),''
JHEP {\bf 9906}, 017 (1999)
[hep-ph/9905351].
%%CITATION = HEP-PH 9905351;%%


%%%%%%%%%%%%%%%%%%%%%%%%%%%%%%%%%%%%%%%%%%%%%%%%%%%%%%%%%%

\end{thebibliography}
\end{document}